\documentclass[aps,prd,twocolumn,floatfix,noshowpacs,tightenlines,noshowkeys,superscriptaddress,amsmath,amssymb,nofootinbib]{revtex4-1}
\usepackage{amssymb,amsbsy,epsfig,color,graphicx}
\usepackage{color}
\usepackage{[longtable}
\usepackage{array}
\usepackage{dcolumn}   % needed for some tables
\usepackage{cellspace}
\usepackage{mathtools}
\usepackage{amstext}
\usepackage{amssymb}
\usepackage{stmaryrd}
\usepackage{stackrel}
\usepackage{graphicx}
\usepackage{esint}
\usepackage[utf8]{inputenc}
\usepackage{adjustbox}
\usepackage{multirow}
\usepackage{float}
\usepackage{booktabs}
\usepackage{enumitem}
\usepackage{slashed}
\usepackage{hyperref}
\usepackage{etoolbox} % for \appto
\usepackage{lipsum} % for mock text
\usepackage[capitalize]{cleveref}
\usepackage{lipsum,booktabs}
\usepackage{changepage}
\usepackage{multirow}
\usepackage[caption=false]{subfig}
\allowdisplaybreaks
\usepackage{soul}
\def\bea{\begin{eqnarray}}
	\def\eea{\end{eqnarray}}

\newcommand{\ba}{\begin{eqnarray}}
	\newcommand{\ea}{\end{eqnarray}}

\makeatletter

\appto{\appendix}{%
	\@ifstar{\def\theequation@prefix{A.}}%
	{}%
}
\makeatother

\hypersetup{pdfstartview=FitV,colorlinks=true,linkcolor=midblue,citecolor=midblue,filecolor=midblue,urlcolor=midblue}

\definecolor{midblue}{rgb}{0,0,0.5}
\definecolor{cadmiumorange}{rgb}{0.93, 0.53, 0.18}
\def\Schld{Schwarzchild}

\def\dS{de Sitter }

\def\mA{\mathcal{A}}
\def\rth{r_{\textrm{th}}}

\newcommand{\mR}{\mathcal{R}}

\usepackage{hyperref}
\hypersetup{colorlinks=true}

\def\equationautorefname~#1\null{%
	Eq.~(#1)\null
}
\def\figureautorefname~#1\null{%
	Fig.~#1\null
}
\def\tableautorefname~#1\null{%
	Table.~#1\null
}
\def\sectionautorefname~#1\null{%
	Section #1\null
}
\def\appendixautorefname~#1\null{%
	Appendix #1\null
}

\begin{document}

	\title{Traversable wormholes in bi-metric gravity}

	\author{Mostafizur Rahman}
	\email{mostafizur.r@iitgn.ac.in}
	\affiliation{Indian Institute of Technology, Gandhinagar, Gujarat-382355, India}
	
	\author{Anjan A Sen}
	\email{aasen@jmi.ac.in}
	\affiliation{Center For Theoretical Physics, Jamia Millia Islamia, New Delhi 110025, India}

        \author{Sunil Singh Bohra}
	\email{ Sunilsinghbohra87@gmail.com}
	\affiliation{Center For Theoretical Physics, Jamia Millia Islamia, New Delhi 110025, India}

	\begin{abstract}
The ghost-free bi-metric gravity theory is a viable theory of gravity that explores the interaction between a massless and a massive graviton and can be  described in terms of two dynamical metrics. In this paper, we present an exact static, spherically symmetric vacuum solution within this theory. The solution is spatially Schwarzschild-de Sitter, with the value of the cosmological constant determined by the graviton mass and the interaction parameters of the theory. Notably, for specific parameter ranges, the solution represents a traversable Lorentzian wormhole that violates the weak energy condition near its throat. Furthermore, we have investigated the evolution of scalar and electromagnetic fields in this wormhole spacetime and observed the presence of arbitrarily long-lived quasi-resonant modes in the quasinormal spectrum.
%In this paper, we report the finding of an exact static, spherically symmetric vacuum solution in this theory. The solution is spatially Schwarzschild-de Sitter, with the value of the cosmological constant determined by the graviton mass and the interaction parameters of the theory. Interestingly, for certain parameter ranges, the solution describes a Lorentzian wormhole with a violation of the weak energy condition occurs near its throat. Additionally, we investigate the evolution of scalar and electromagnetic fields in this wormhole spacetime and observe the existence of arbitrarily long-lived modes, so-called the quasi-resonant modes in the quasinormal spectrum.

	\end{abstract}
	
	\maketitle

	%%%%%%%%%%%%%%%%%%%%%%%%%%%%%%%%%%%%%%%%%%%%%%%%%%%%%%%%%%%%%%%%%%%%%%%%
\section{Introduction}
Ever since the observation of Type-Ia Supernovae \citep{Riess_1998,Perlmutter_1999,1998ApJ...507...46S} have confirmed the existence of late time accelerating Universe at the background level, the search for the extra relativistic dark candidate having repulsive gravity has been the major research goal in the field of particle cosmology and astrophysics. Later additional observational results by multiple Baryon Acoustic Oscillations (BAO) measurements \citep{alam} from different galaxy surveys as well as the measurement of temperature fluctuations in Cosmic Microwave Background Radiation (CMBR)\citep{planck_2015_cosmo,planck_2018_cosmo}, have shown that a small positive cosmological constant ($\Lambda$) in the energy budget of the Universe can be a possible explanation for this late time acceleration. Hence the concordance $\Lambda$CDM \cite{2021arXiv210505208P} has been the simplest yet largely successful model in explaining the observable Universe. The only question that remains is how to achieve the tiny value for the observed $\Lambda$ in an acceptable theoretical construction of quantum field theory which in general always come out with a $\Lambda$ which is many order of magnitude larger than the observed value.

But in recent times few significant observational discrepancies related to $\Lambda$CDM model have put a big question mark on the acceptability of $\Lambda$CDM as the viable model for our observed Universe. Issues like Hubble Tension \cite{SH0ES_2016ApJ...826...56R}, $S_{8}/\sigma_{8}$ tension \citep{Joudaki_2016,Asgari_2021,Troxel_2018}, the observed high-density massive galaxies at very redshifts by JWST \citep{Boylan_Kolchin_2023} and others have started giving undeniable hints that possibly we need to consider model beyond $\Lambda$CDM which is not only consistent with host of observations from SnIa, BAO measurements, CMBR measurements, growth measurements, but also results observed values of cosmological parameters which are consistent across different observations.

This motivates the renewed interest in looking for different modifications in the $\Lambda$CDM model, both at early times as well as in late times. These include approaches involving modification of the gravity at large cosmological scales. Considering the modified gravity models as a possible alternative to $\Lambda$CDM, we should ensure that such modifications restore General Relativity (GR) on small scales. This is needed to match with the local observations like Solar System tests \cite{Will_2014}. The first attempt to modify GR by introducing mass to the intermediate particle for the gravitational force, the graviton, through linear theory of massive gravity was done by Fierz and Pauli  \cite{1939RSPSA.173..211F} . But this theory contains Boulware-Desert (BD) Ghost  \cite{1972PhRvD...6.3368B}. This BD ghost can be removed by inclusion of a second metric $f_{\mu\nu}$ into the theory alongside the mass-less metric $g_{\mu\nu}$ with carefully constructed interaction term between these two metrics \cite{2011PhRvL.106w1101D}. Dynamics of the second metric give rise to the bimetric gravity  \cite{2012JHEP...02..126H}. The bimetric gravity has a screening mechanism that can restore the general relativity on solar system scales \cite{Babichev_2013}. 

The quest for finding stationary black hole solutions within bi-metric gravity theory has a long and rich history. The first static, spherically symmetric black hole solution for a specific class of bi-metric gravity theory was found nearly four decades ago \cite{PhysRevD.18.1047, PhysRevD.20.1019, Gurses:1981an}. The metric functions $g_{\mu\nu}$ and $f_{\mu\nu}$ in this solution are not bi-diagonalizable and, when expressed in Eddington-Finkelstein coordinates, they belong to the family of \Schld-de Sitter black holes \cite{PhysRevD.89.081502, Volkov:2014ooa}. Subsequently, charged and rotating counterparts of these solutions were also found \cite{Babichev:2014fka, Babichev:2014tfa}.  Interestingly, it was noticed that they theory only admits \Schld-de Sitter (or its rotating counterpart) family of solutions when the two metric functions are proportional to each other i.e., when $g_{\mu\nu}=C^2 f_{\mu\nu}$, where $C$ is some constant \cite{Volkov:2014ooa}. The theory also admits hairy black hole solutions \cite{1989PZETF..50..312V,Volkov:2014ooa, Volkov:1998cc, Volkov:2016ehx, Gervalle_2020, Volkov_2012, PhysRevD.83.084042, Berezhiani:2008nr}. However, less emphasis was placed on finding static, spherically symmetric solutions that are not black holes. It is only recently that Sushkov and Volkov numerically obtained wormhole solutions within the framework of bi-metric gravity \citep{Sushkov_2015}. \par

%Regarding the Schwarzschild-like solution in bimteric gravity, there are static spherically symmetric black hole solutions in bimetric gravity which can be both asymptotically flat \cite{Gervalle_2020} and asymptotically non-flat \cite{Volkov_2012}. Sushkov and Volkov have obtained numerically wormhole solutions in bimetric gravity \citep{Sushkov_2015}. 

In this work, we obtain an analytical, closed-form wormhole solution in bimteric gravity. We also study the scalar and vector perturbations in the wormhole spacetime and show that they are stable.\par
We organize the paper in the following manner: In Section \ref{Sec:Solution}, we briefly describe the bi-metric gravity theory and present a new exact static, spherically symmetric solution within this theory. In Section \ref{Sec:Wormhole}, we show that the solution describes a wormhole for a certain parameter range and check whether the spacetime respects the energy conditions. In Section \ref{Sec:Perurbation}, we study the perturbation of the wormhole spacetime by scalar and electromagnetic field. We show the existence of arbitrarily long-lived modes in the quasinormal spectrum through both analytical and numerical methods in Section \ref{Sec:QRM}. The conclusion of the paper is presented in Section \ref{Sec:Conclusion}. In Appendix \ref{App:Bi-diagonal}, we investigate the properties of solutions within a specific class of bi-gravity theory. Lastly, in Appendix \ref{App:Circular}, we present a discussion on the roots of a cubic equation.\par

\textit{Notation and Conventions:} Throughout the paper, we adopt positive sign conventions $(-1,1,1,1)$ for both the metrics and set the fundamental constants as $G=c=\hbar=1$.

\section{Static and Spherically symmetric solutions in Bi-metric gravity theory}\label{Sec:Solution}
The ghost-free bi-gravity theory describes the interaction between two gravitons, with one being massless and the other being massive. In this theory, the spacetime is characterized by two metric functions, namely $g_{\mu\nu}$ and $f_{\mu\nu}$. The kinetic term in the action for both metrics follows the standard Einstein-Hilbert action while a local potential term governs the interaction between the gravitons, which does not contain any derivatives of the metric functions. The action of this theory is given by \cite{Hassan:2011vm} 
%%%%%%%%%%%%%%%%%%%%%%%%%%%%%%%%%%%%%%%%%%%%%%%%%%%%%%%%%%%%%%%%%%%%%%%%%%%%%%%%%%%%%%%%%%%%%%%%%%%
\begin{equation}\label{bi_gravity_action}
    \begin{aligned}
    S&=-\frac{M_g^2}{2}\int d^4x\sqrt{-g} R-\frac{M_f^2}{2}\int d^4x\sqrt{-f}\mathcal{R}\\&+m^2M_g^2\int d^4 x\sqrt{-g}\sum_{n=0}^{4}\beta_n e_n(\chi)+\int d^4x\sqrt{-g}\mathcal{L}_m
    \end{aligned}
\end{equation}
%%%%%%%%%%%%%%%%%%%%%%%%%%%%%%%%%%%%%%%%%%%%%%%%%%%%%%%%%%%%%%%%%%%%%%%%%%%%%%%%%%%%%%%%%%%%%%%%%%%
where, $M_g$ and $M_f$ are the Planck masses corresponding to the metric $g_{\mu\nu}$ and $f_{\mu\nu}$. $m$ is the mass of the graviton. $R_g$ and $\mR$ are the Ricci scalar for $g_{\mu\nu}$ and $f_{\mu\nu}$ respectively, $\chi=\sqrt{g^{-1}f}$ is a matrix defined in such way that $\chi^2=g_{\mu\nu}f^{\mu\nu}$. $e_n(\chi)$ is the elementary symmetry polynomials of eigen value the matrix $\chi$ which can be written as follows\cite{Hassan:2012wr} 
%%%%%%%%%%%%%%%%%%%%%%%%%%%%%%%%%%%%%%%%%%%%%%%%%%%%%%%%%%%%%%%%%%%%%%%%%%%%%%%%%%%%%%%%%%%%%%%%%%%
\begin{equation}\label{e_n}
    \begin{aligned}
    e_0(\chi)&=1,\quad e_1(\chi)=[\chi],\quad e_2(\chi)=\frac{1}{2}\left([\chi]^2-[\chi^2]\right)\\
    e_3(\chi)&=\frac{1}{6}\left([\chi]^3-3[\chi][\chi^2]+2[\chi^3]\right), \quad e_4(\chi)=\textrm{det}(\chi)
    \end{aligned}
\end{equation}
%%%%%%%%%%%%%%%%%%%%%%%%%%%%%%%%%%%%%%%%%%%%%%%%%%%%%%%%%%%%%%%%%%%%%%%%%%%%%%%%%%%%%%%%%%%%%%%%%%%
where, $[\chi]$ is the trace of the matrix $\chi$ and $\textrm{det}(\chi)$ is the determinant of $\chi$. $\beta_n$'s are free parameters. The equation of motion is obtained by varying the action with respect to $g_{\mu\nu}$ and $f_{\mu\nu}$, which can be expressed as follows \citep{Hassan:2012wr}
%%%%%%%%%%%%%%%%%%%%%%%%%%%%%%%%%%%%%%%%%%%%%%%%%%%%%%%%%%%%%%%%%%%%%%%%%%%%%%%%%%%%%%%%%%%%%%%%%%%
\begin{equation}\label{einstein_equation}
    \begin{aligned}
     R_{\mu\nu}-\frac{1}{2}g_{\mu\nu}R &=T_{\mu\nu}^{g}+\frac{1}{M_g^2}T_{\mu\nu}\\
     \mR_{\mu\nu}-\frac{1}{2}f_{\mu\nu}\mR &=T_{\mu\nu}^{f}
    \end{aligned}
\end{equation}
%%%%%%%%%%%%%%%%%%%%%%%%%%%%%%%%%%%%%%%%%%%%%%%%%%%%%%%%%%%%%%%%%%%%%%%%%%%%%%%%%%%%%%%%%%%%%%%%%%%
where, $M_{*}=M_f/M_g$, and 
%%%%%%%%%%%%%%%%%%%%%%%%%%%%%%%%%%%%%%%%%%%%%%%%%%%%%%%%%%%%%%%%%%%%%%%%%%%%%%%%%%%%%%%%%%%%%%%%%%%
\begin{equation}\label{effective_eom}
    \begin{aligned}
T_{\mu\nu}^{g} =-\frac{m^2}{2}\sum_{n=0}^{3}(-1)^n \beta_n\bigg[g_{\mu\lambda}&Y_{(n)\nu}^{\lambda}(\chi)+g_{\nu\lambda}Y_{(n)\mu}^{\lambda}(\chi)\bigg]\\
 T_{\mu\nu}^{f}= -\frac{m^2}{2M_{*}^2}\sum_{n=0}^{3}(-1)^n \beta_{4-n}\bigg[&f_{\mu\lambda}Y_{(n)\nu}^{\lambda}(\chi^{-1})\\&+f_{\nu\lambda}Y_{(n)\mu}^{\lambda}(\chi^{-1})\bigg]
\end{aligned}
\end{equation}
%%%%%%%%%%%%%%%%%%%%%%%%%%%%%%%%%%%%%%%%%%%%%%%%%%%%%%%%%%%%%%%%%%%%%%%%%%%%%%%%%%%%%%%%%%%%%%%%%%%
with
%%%%%%%%%%%%%%%%%%%%%%%%%%%%%%%%%%%%%%%%%%%%%%%%%%%%%%%%%%%%%%%%%%%%%%%%%%%%%%%%%%%%%%%%%%%%%%%%%%%
\begin{equation}\label{Y_n}
    \begin{aligned}
    Y_{(0)}(\chi)&=1,\quad Y_{(1)}(\chi)=\chi-\mathbb{I} [\chi],\\Y_{(2)}(\chi)&=\chi^2-\chi[\chi]+\frac{1}{2}\mathbb{I}\left([\chi]^2-[\chi^2]\right)\\
    Y_{(3)}(\chi)&=\chi^{3}-\chi^{2}[\chi]+\frac{1}{2}\chi \left([\chi]^2-[\chi^2]\right)\\&-\frac{1}{6}\left([\chi]^3-3[\chi][\chi^2]+2[\chi^3]\right)
    \end{aligned}
\end{equation}
%%%%%%%%%%%%%%%%%%%%%%%%%%%%%%%%%%%%%%%%%%%%%%%%%%%%%%%%%%%%%%%%%%%%%%%%%%%%%%%%%%%%%%%%%%%%%%%%%%%
Under the scaling transformation $f_{\mu\nu}\to \frac{M_g^2}{M_f^2}f_{\mu\nu}$ and $\beta_n\to \left(\frac{M_f}{M_g}\right)^n \beta_n$, we can make $M_{*}^2=1$. Hence it is not a free parameter. In what follows, we consider $M_{*}^2=1$ and $M_g=M_{\textrm{Pl}}$. As a consequence of the Bianchi identity and covariant conservation equation of the energy momentum tensor, we have the following relation
%%%%%%%%%%%%%%%%%%%%%%%%%%%%%%%%%%%%%%%%%%%%%%%%%%%%%%%%%%%%%%%%%%%%%%%%%%%%%%%%%%%%%%%%%%%%%%%%%%%
    \begin{align}
    &\nabla^{\mu}\sum_{n=0}^{3}(-1)^n \beta_n\left[g_{\mu\lambda}Y_{(n)\nu}^{\lambda}(\chi)+g_{\nu\lambda}Y_{(n)\mu}^{\lambda}(\chi)\right]=0~,\label{BIi}\\
    & \bar{\nabla}^{\mu}\sum_{n=0}^{3}(-1)^n \beta_{4-n}\left[f_{\mu\lambda}Y_{(n)\nu}^{\lambda}(\chi^{-1})+f_{\nu\lambda}Y_{(n)\mu}^{\lambda}(\chi^{-1})\right]=0~,\label{BIii}
    \end{align}
%%%%%%%%%%%%%%%%%%%%%%%%%%%%%%%%%%%%%%%%%%%%%%%%%%%%%%%%%%%%%%%%%%%%%%%%%%%%%%%%%%%%%%%%%%%%%%%%%%%
where, $\nabla_{\mu}$ and $\bar{\nabla}_{\mu}$ are the covariant derivative with respect to the metric $g_{\mu\nu}$ and $f_{\mu\nu}$ respectively. However, \autoref{BIi} and \autoref{BIii} are equivalent.\par
In this paper, we focus on a specific class of bi-gravity theory with $\beta_{2}=\beta_{3}=0$. Our objective is to find a solution where both metrics are static, spherically symmetric, and diagonal in the spherical coordinate $(t,~r,~\theta,~\phi)$. Wormhole solutions in bi-gravity theories under the bidiagonal assumption has previously been studied in \cite{Sushkov_2015}. Also \cite{Schmidt_May_2016} discussed conditions for the bidiagonal assumption for the metrics. We should also refer interested readers to \cite{Torsello_2017} (and references therein) for a detailed discussion on this issue. Following \cite{Schmidt_May_2016}, we explicitly show in Appendix \ref{App:Bi-diagonal} that the static, spherically symmetric, vacuum solutions of the bi-gravity theory are bi-diagonal under the condition $\beta_2=\beta_3=0$.
%As demonstrated in \cite{Schmidt_May_2016}, with our assumption $\beta_{2}=\beta_{3}=0$, there is no  nonbidiagonal solutions for the field equations. We have
With this, we consider the following ansatz for the metrics
\begin{equation}
 ds_{g}^{2}  =-e^{2 a}dt^{2}+e^{2 b} dr^{2}+r^{2}\left(d\theta^{2}+\sin^{2} \theta d \phi^{2}\right)
\end{equation}
and
\begin{equation}
    ds_{f}^{2} =-e^{2 A} d t^{2}+e^{2 B} d r^{2}+r^{2}\left(d \theta^{2}+\sin ^{2} \theta d \phi^{2}\right)
\end{equation}
where $a, b, A, B$ are functions of $r$ only. We are interested in the vacuum solution($T_{\mu\nu}=0$) outside a spherical source.

\noindent For $g_{\mu\nu}$ metric, `$tt$' and `$rr$' components of field equations (\autoref{einstein_equation}) are,

\begin{equation}\frac{1}{r^{2}}+e^{-2b}\left(\frac{2 b^{\prime}}{r}-\frac{1}{r^{2}}\right)-m^{2}\left[\beta_{0}+\beta_{1}\left(2+e^{B-b}\right)\right]=0~,\label{EFEtt}\end{equation}
and
\begin{equation}-\frac{1}{r^{2}}+e^{-2b}\left(\frac{2 a^{\prime}}{r}+\frac{1}{r^{2}}\right)+m^{2}\left[\beta_{0}+\beta_{1}\left(2+e^{A-a}\right)\right]=0~,\label{EFErr}\end{equation}
respectively.

For $f_{\mu\nu}$ metric, `$tt$' and `$rr$' components of field equations (\autoref{einstein_equation}) are

\begin{equation}\frac{1}{r^{2}}+e^{-2B}\left(\frac{2B^{\prime}}{r}-\frac{1}{r^{2}}\right)-\beta_{1} m^{2} e^{b-B} + \beta_{4} m^{2}=0~,\label{EFEAtt}\end{equation}
and
\begin{equation}-\frac{1}{r^{2}}+e^{-2B}\left(\frac{2 A^{\prime}}{r}+\frac{1}{r^{2}}\right)+\beta_{1} m^{2} e^{a-A} -\beta_{4} m^{2} = 0~,\label{EFEArr}\end{equation}
respectively.

These are the independent equations. Other remaining equations can be obtained with the help of these given equations and Bianchi identity. There are two Bianchi identities for $g_{\mu\nu}$ and $f_{\mu\nu}$ given by \autoref{BIi} and \autoref{BIii} but as mentioned earlier, they are equivalent and the resulting equation is given by

\begin{equation}e^{(a-A)}\left[a^{\prime}+\frac{2}{r}\right]-A^{\prime}e^{(b-B)}-\frac{2}{r}e^{(a-A)}e^{(b-B)} = 0
\label{bianchi}
\end{equation}

\noindent
To solve the system of equations, we assume the following condition: $g_{rr}=f_{rr}$ which gives $b= B$. This reduces \autoref{bianchi}  as a first integral:

\begin{equation}e^{a}-e^{A} = 2\kappa
\label{bianchi_rel}
\end{equation},

\noindent
where $\kappa$ is the constant of integration. All our subsequent calculations and results are under the above mentioned assumption.

By solving (\autoref{EFEAtt}.), we get;
\begin{equation}
\left(g_{11}\right)^{-1} = e^{-2B}=1-\frac{2M}{r}-\frac{(\beta_{1}-\beta_{4})m^{2}r^{2}}
 {3},
 \end{equation}
where $M$ is the constant of integration. Adding (\autoref{EFEtt}) and (\autoref{EFErr}), we get

\begin{equation}\label{17}
\frac{2}{r} e^{-2b}\left[a^{\prime}+b^{\prime}\right]+m^{2} \beta_{1}\left(e^{A-a}-e^{B-b}\right) = 0,
\end{equation}
and adding (\autoref{EFEAtt}) and (\autoref{EFEArr}), we get
\begin{equation}\label{18}
\frac{2}{r} e^{-2B}\left(A^{\prime}+B^{\prime}\right)+m^{2} \beta_{1}\left(e^{a-A}-e^{b-B}\right)=0.
\end{equation}

Applying our assumption $b=B$, and the relation is given by \autoref{bianchi_rel}, \autoref{17} and \autoref{18} become

\begin{equation}\label{EFE}
\frac{2}{r} e^{-2b}e^{a}\left(a^{\prime}+b^{\prime}\right)-2m^{2}\beta_{1}\kappa = 0~,
\end{equation}

\begin{equation}\frac{2}{r}e^{-2b}\left[a^{\prime}e^{a}+b^{\prime}\left(e^{a}-2\kappa\right)\right]+2m^{2}\beta_{1}\kappa=0~,\label{EFEA}
\end{equation}
which on adding gives
\begin{equation}\frac{\left(e^{a}\right)^{\prime}}{\left(e^{a}-\kappa\right)}+b^{\prime} = 0.
\end{equation}
On Integrating the above equation, we get
\begin{equation}e^{a}=\kappa+\lambda e^{-b},
\end{equation}
where $\exp{[c_2]}= \lambda$ is the constant of integration. With this the metric components of $g_{\mu\nu}$ and $f_{\mu\nu}$ become:
\begin{equation}\label{metricg}
g_{00}=e^{2a} = \left(\kappa+\lambda \sqrt{1-\frac{2M}{r}-\frac{(\beta_{1}-\beta_{4})m^{2}r^{2}}{3}}\right)^{2},\end{equation}
\begin{equation}\label{metricf}
f_{00} = e^{2b} =\left(\kappa-\lambda \sqrt{1-\frac{2M}{r}-\frac{(\beta_{1}-\beta_{4})m^{2}r^{2}}{3}}\right)^{2},\end{equation}  
For the sake of completeness, here we also present the $g_{11}$ and $f_{11}$ components of the metrics
\begin{equation}
\left(g_{11}\right)^{-1}=\left(f_{11}\right)^{-1} = 1-\frac{2M}{r}-\frac{(\beta_{1}-\beta_{4})m^{2}r^{2}}{3}.
\end{equation}

 On our given assumption, \autoref{EFEtt} and \autoref{EFEAtt} are decoupled. Both of them can now easily integrated. One gives a solution in the term of parameter $\beta_{0}$ and $\beta_{1}$, While other gives in term of parameter $\beta_{1}$ and $\beta_{4}$. By equating both the solution, we see that, the non zero parameters $\beta_{0}, \beta_{1}$ and $\beta_{4}$ are not independent and they are related by the consistency relation: 
$$\beta_{0} + 2\beta_{1} + \beta_{4} = 0$$
Before investigating the properties of the above mentioned solution, it is crucial to verify its compatibility with the theorem put forth by Deffayet and Jacobson for the bi-gravity theory \cite{Deffayet:2011rh}. The theorem asserts that if the two metrics $g_{\mu\nu}$ and $f_{\mu\nu}$ are static, spherically symmetric, diagonal, and non-singular, then their Killing horizons (if they exists!) should coincide. We examine whether this condition holds true for our proposed solution. Since the solutions are static, the Killing horizons of the metrics can be found by solving the equations $g_{00}=0$ and $f_{00}=0$, respectively. From the equations \autoref{metricg} and \autoref{metricf}, it can be verified that the Killing horizon of both metrics is a solution of the following equation
%%%%%%%%%%%%%%%%%%%%%%%%%%%%%%%%%%%%%%%%%%%%%%%%%%%%%%%%%%%%%%%%%
\begin{equation}\label{Decartes1}
\begin{aligned}
r^3+\frac{3}{\Lambda}\left(\frac{\kappa^2}{\lambda^2}-1\right)r+\frac{6M}{\Lambda}=0
\end{aligned}
\end{equation}
%%%%%%%%%%%%%%%%%%%%%%%%%%%%%%%%%%%%%%%%%%%%%%%%%%%%%%%%%%%%%%%%%
Hence, the Killing horizon of both metrics coincides with each other. Having established that the solution is compatible with Deffayet and Jacobson's theorem, we will further investigate the properties of the solution in the next section.
%%%%%%%%%%%%%%%%%%%%%%%%%%%%%%%%%%%%%%%%%%%%%%%%%%%%%%%%%%%%%%%%%
%%%%%%%%%%%%%%%%%%%%%%%%%%%%%%%%%%%%%%%%%%%%%%%%%%%%%%%%%%%%%%%%%
\section{Lorentzian wormholes in bi-metric gravity}\label{Sec:Wormhole}
The general, static and spherically symmetric solution of \autoref{einstein_equation} with the assumption $\beta_2=\beta_3=0$ is given by the line element 
%%%%%%%%%%%%%%%%%%%%%%%%%%%%%%%%%%%%%%%%%%%%%%%%%%%%%%%%%%%%%%%%%
\begin{equation}\label{bi-gravity_wormhole}
\begin{aligned}
   ds_g^2&=- \left(\kappa+\lambda \sqrt{1-\frac{2M}{r}-\frac{\Lambda r^{2}}{3}}\right)^{2} dt^2\\&+\left(1-\frac{2M}{r}-\frac{\Lambda r^{2}}{3}\right)^{-1}dr^2+r^{2}\left(d \theta^{2}+\sin ^{2} \theta d \phi^{2}\right)  
\end{aligned}
\end{equation}
%%%%%%%%%%%%%%%%%%%%%%%%%%%%%%%%%%%%%%%%%%%%%%%%%%%%%%%%%%%%%%%%%
 where, $\Lambda=(\beta_{1}-\beta_{4})m^{2}$. In what follows, we consider $\Lambda>0$ i.e., $\beta_1>\beta_4$. From \autoref{einstein_equation}, we can see that the metric $f_{\mu\nu}$ does not couple to the matter field and thus has no direct observational significance. Therefore, we only concern ourself with the metric $g_{\mu\nu}$.\par
 Note that the metric \autoref{bi-gravity_wormhole} corresponds to \Schld-\dS black hole \cite{PhysRevD.15.2738} for $\kappa=0$ and $\lambda=1$. In that scenario, the position of the horizons corresponds to the solution of the following cubic equation
 %%%%%%%%%%%%%%%%%%%%%%%%%%%%%%%%%%%%%%%%%%%%%%%%%%%%%%%%%%%%%%%%%
\begin{equation}\label{throat}
\begin{aligned}
1-\frac{2M}{r}-\frac{\Lambda r^{2}}{3}=0 
\end{aligned}
\end{equation}
%%%%%%%%%%%%%%%%%%%%%%%%%%%%%%%%%%%%%%%%%%%%%%%%%%%%%%%%%%%%%%%%%
The above equation has two positive roots $r_c$ and $r_{h}$ ($r_c\geq r_h$) corresponding to its cosmological and event horizon, respectively, and a negative root $r_0=-(r_c+r_h)$ if the following condition is satisfied: $0<\Lambda\leq 1/9M^2$ (see Appendix \ref{App:Circular} for a elaborate discussion on the solution of a cubic equation).\par
However, \autoref{bi-gravity_wormhole} describes a wormhole geometry if the following condition is satisfied \cite{10.1119/1.15620, Visser:1995cc, Lemos:2003jb}:
\begin{enumerate}
    \item[\textbf{ I.}] The geometry does not have a horizon i.e., $-g_{00}=\exp{(2a)}$ has no real, positive roots. Here, $a(r)$ is called the \textit{redshift} function.
    \item[\textbf{II.}] The geometry has a minimum radius $\rth$, known as the \textit{throat} of the wormhole which is the smallest positive solution of $g^{11}=0$.
\end{enumerate}
%%%%%%%%%%%%%%%%%%%%%%%%%%%%%%%%%%%%%%%%%%%%%%%%%%%%%%%%%%%%%%%%%%%%%%%%%%%%%%%%%%%%%%%%%%%%%%%%%%%%%
The simplest choice to ensure that the no-horizon condition (Condition \textbf{I}) is satisfied ( thus allowing a wormhole solution) is given by $\kappa\neq 0$, $\lambda=0$. This choice corresponds to a spatial-\Schld-\dS traversable wormhole. For non-vanishing $\lambda$, we have to ensure that $g_{00}=0$ does not have positive real roots. We can rewrite this equation in the following manner 
%%%%%%%%%%%%%%%%%%%%%%%%%%%%%%%%%%%%%%%%%%%%%%%%%%%%%%%%%%%%%%%%%
\begin{equation}\label{Decartes}
\begin{aligned}
r^3+\frac{3}{\Lambda}\left(\frac{\kappa^2}{\lambda^2}-1\right)r+\frac{6M}{\Lambda}=0
\end{aligned}
\end{equation}
%%%%%%%%%%%%%%%%%%%%%%%%%%%%%%%%%%%%%%%%%%%%%%%%%%%%%%%%%%%%%%%%%
Using Descartes' rule of sign \cite{doi:10.1080/00029890.1998.12004907}, we can show that the equation has no positive roots (only one negative root) when the following condition is satisfied $\kappa^2\geq\lambda^2$. This choice restricts the appearance of the event horizon and, thus, gives us a wormhole solution.\par 
To get a better understanding of Condition II, we make use of embedding diagrams \cite{10.1119/1.15620, Visser:1995cc, Lemos:2003jb}. Since the spacetime is spherically symmetric, we can confine our attention to equatorial plane $\theta=\pi/2$ without the loss of generality.  Under such consideration, the  $t=\textrm{const}$ hypersurface becomes
%%%%%%%%%%%%%%%%%%%%%%%%%%%%%%%%%%%%%%%%%%%%%%%%%%%%%%%%%%%%%%%%%
\begin{equation}\label{2dslice}
ds_g^2=g_{11} dr^2+r^2 d\phi^2
\end{equation}
%%%%%%%%%%%%%%%%%%%%%%%%%%%%%%%%%%%%%%%%%%%%%%%%%%%%%%%%%%%%%%%%%
%%%%%%%%%%%%%%%%%%%%%%%%%%%%%%%%%%%%%%%%%%%%%%%%%%%%%%%%%%%%%%%%%%%%%%%%%%%%%%%%%%%%%%%%%%%%%%%%%%%
\begin{figure*}[htb!]
	%%%%%%%%%%%%%%%%%%%%%%%%
	\centering
	\minipage{0.48\textwidth}
	\includegraphics[width=\linewidth]{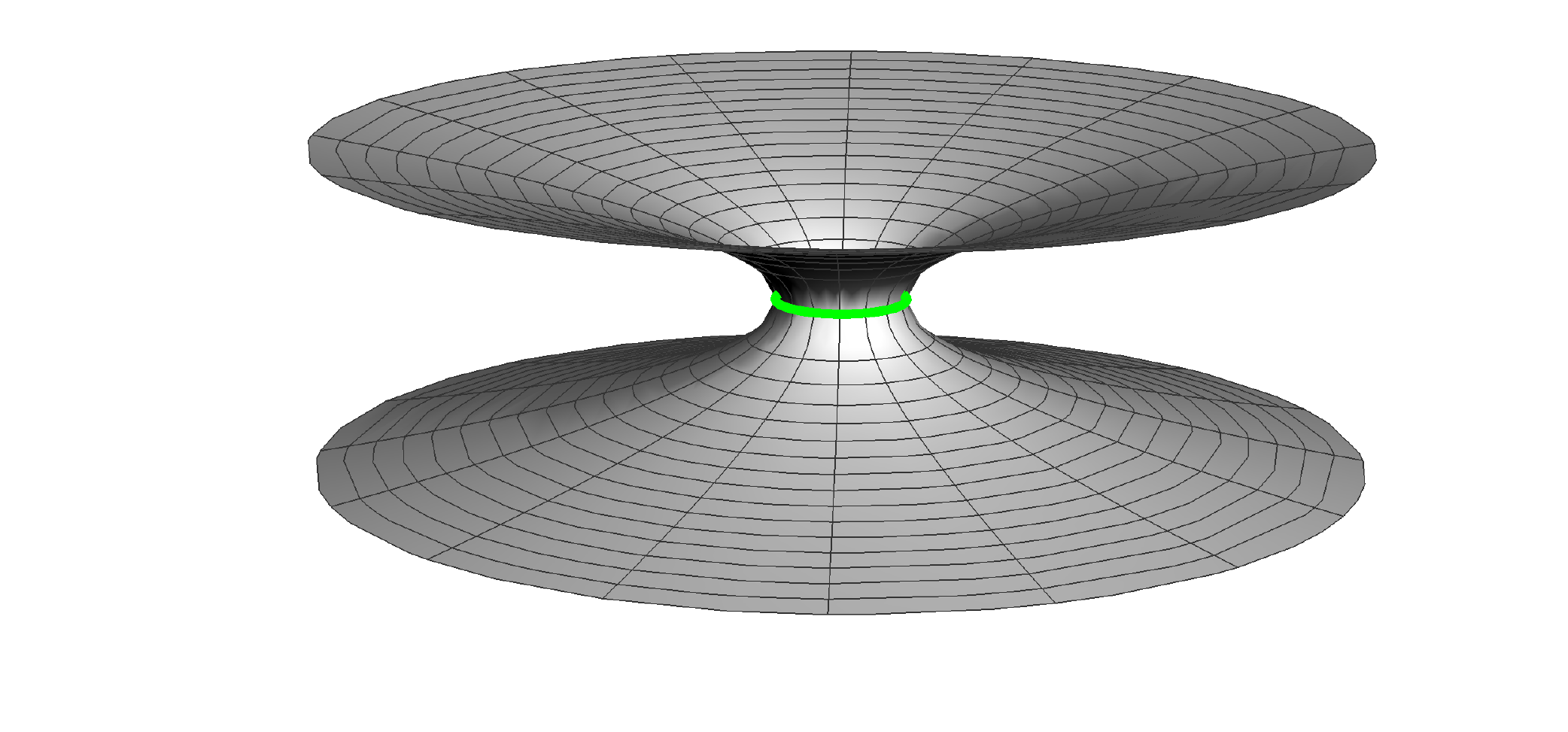}
% \caption{Wormholes for $\Lambda=0$}
	\endminipage\hfill
	%%%%%%%%%%%%%%%%%%%%%%%%
	\minipage{0.48\textwidth}
	\includegraphics[width=\linewidth]{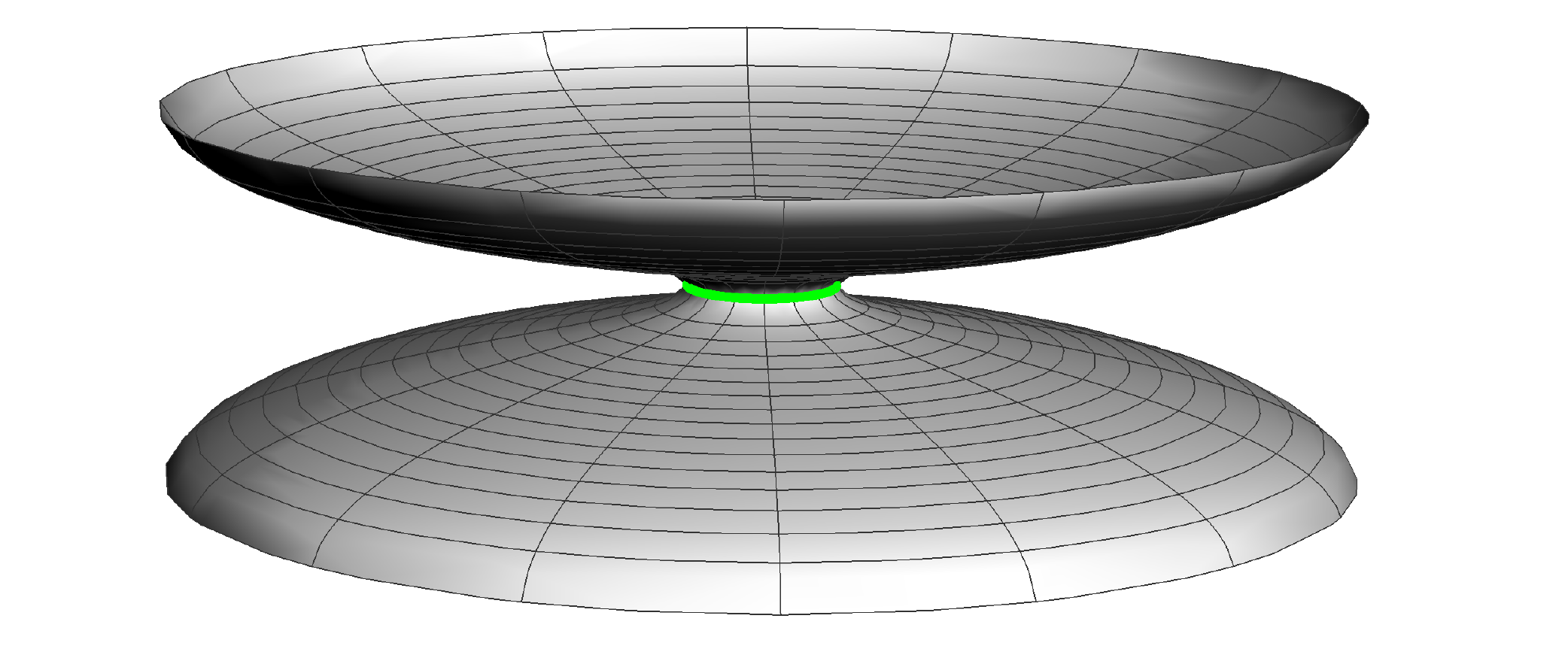}
% \caption{Wormholes for $\Lambda>0$}
	\endminipage
	\caption{The embedding diagram for a wormhole with the metric \autoref{bi-gravity_wormhole}. The left panel depicts a wormhole with vanishing $\Lambda$ whereas the right panel depicts a wormhole with $\Lambda>0$. The wormhole throat, shown as a green ring, connects two distinct universes.}\label{fig_wormholes}
\end{figure*}	
%%%%%%%%%%%%%%%%%%%%%%%%%%%%%%%%%%%%%%%%%%%%%%%%%%%%%%%%%%%%%%%%%
We embed this two-dimensional slice into a three-dimensional Cylindrical spacetime  $(z,r,\phi)$
%%%%%%%%%%%%%%%%%%%%%%%%%%%%%%%%%%%%%%%%%%%%%%%%%%%%%%%%%%%%%%%%%
\begin{equation}\label{3dslice}
\begin{aligned}
ds_g^2&=dz^2+ dr^2+r^2 d\phi^2\\
&=\left[1+\left(\dfrac{dz}{dr}\right)^2\right]dr^2+r^2 d\phi^2\
\end{aligned}
\end{equation}
%%%%%%%%%%%%%%%%%%%%%%%%%%%%%%%%%%%%%%%%%%%%%%%%%%%%%%%%%%%%%%%%%
Comparing \autoref{2dslice} and \autoref{3dslice}, we find the following equation for the embedding surface  
%%%%%%%%%%%%%%%%%%%%%%%%%%%%%%%%%%%%%%%%%%%%%%%%%%%%%%%%%%%%%%%%%
\begin{equation}\label{embedding}
\begin{aligned}
\left(\dfrac{dz}{dr}\right)^2=g_{11}-1=\frac{1-g^{11}}{g^{11}}
\end{aligned}
\end{equation}
%%%%%%%%%%%%%%%%%%%%%%%%%%%%%%%%%%%%%%%%%%%%%%%%%%%%%%%%%%%%%%%%%
Here, we use the fact that $g^{11}=g_{11}^{-1}$. The location of the wormhole throat is defined as the minimum radius at which the embedding surface becomes vertical, i.e., $dz/dr\to \infty$ \cite{10.1119/1.15620}. By utilizing equation \autoref{embedding}, we can demonstrate that the position of the throat corresponds to the smallest positive solution of $g^{11}=0$. Notably, in the case of \autoref{bi-gravity_wormhole}, the position of the wormhole throat coincides with the location of the horizon of a \Schld-\dS\ black hole. In \autoref{fig_wormholes}, we present the embedding diagram with metric \autoref{bi-gravity_wormhole} for $\Lambda=0$ (left panel) and $\Lambda>0$ (right panel).
	%%%%%%%%%%%%%%%%%%%%%%%%%%%%%%%%%%%%%%%%%%%%%%%%%%%%%%%%%%%%%%%%%%%%%%%%%%%%%%%%%%%%%%%%%%%%%%%%%%%%%
 %%%%%%%%%%%%%%%%%%%%%%%%%%%%%%%%%%%%%%%%%%%%%%%%%%%%%%%%%%%%
%%%%%%%%%%%%%%%%%%%%%%%%%%%%%%%%%%%%%%%%%%%%%%%%%%%%%%%%%%%%%%%%%%%%%%%%%%%%%%%%%%%%%%%%%%%%%%%%%%%
\begin{figure*}[htb!]
	%%%%%%%%%%%%%%%%%%%%%%%%
	\centering
	\minipage{0.33\textwidth}
	\includegraphics[width=\linewidth]{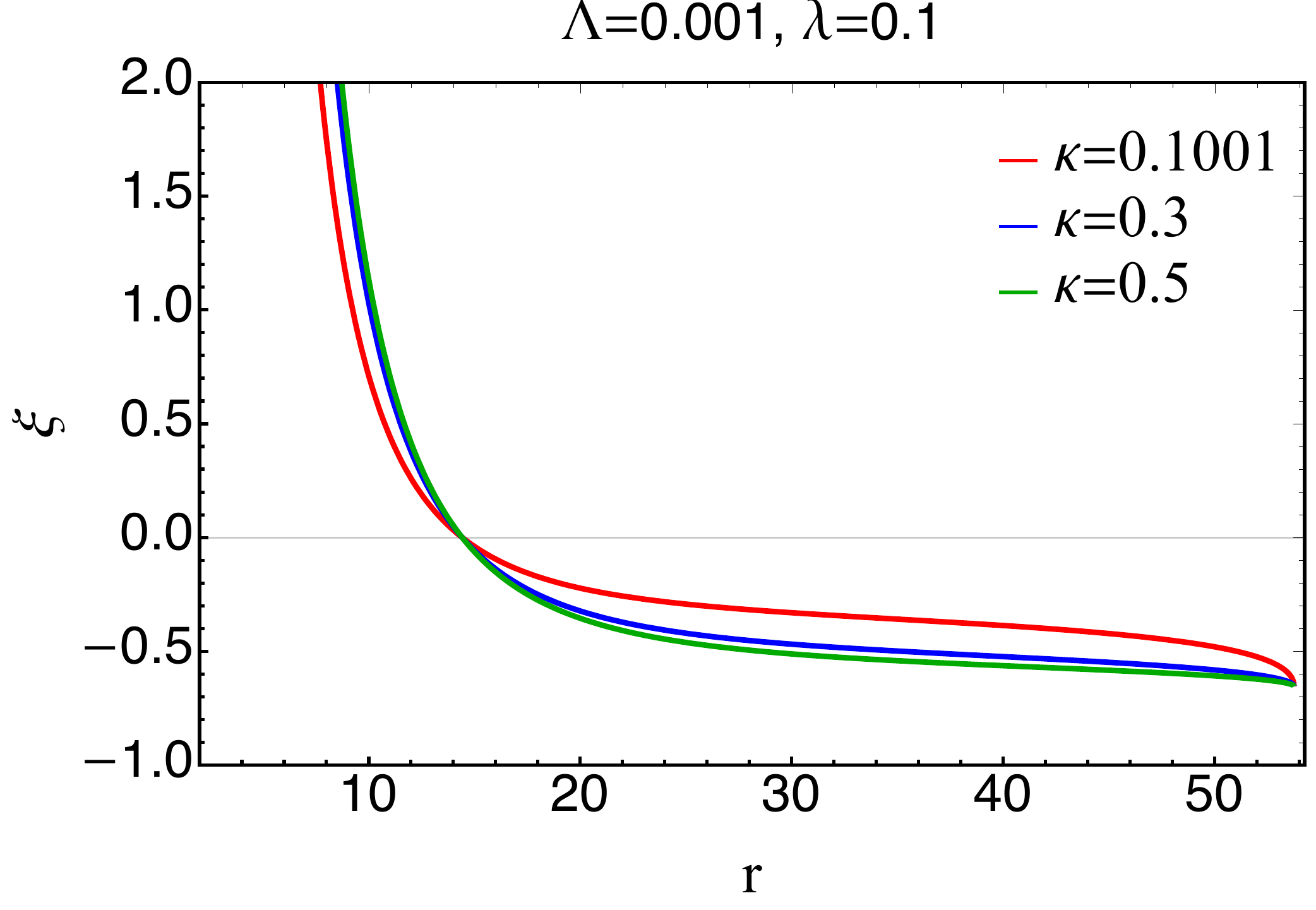}
	\endminipage\hfill
	%%%%%%%%%%%%%%%%%%%%%%%%
	\minipage{0.33\textwidth}
	\includegraphics[width=\linewidth]{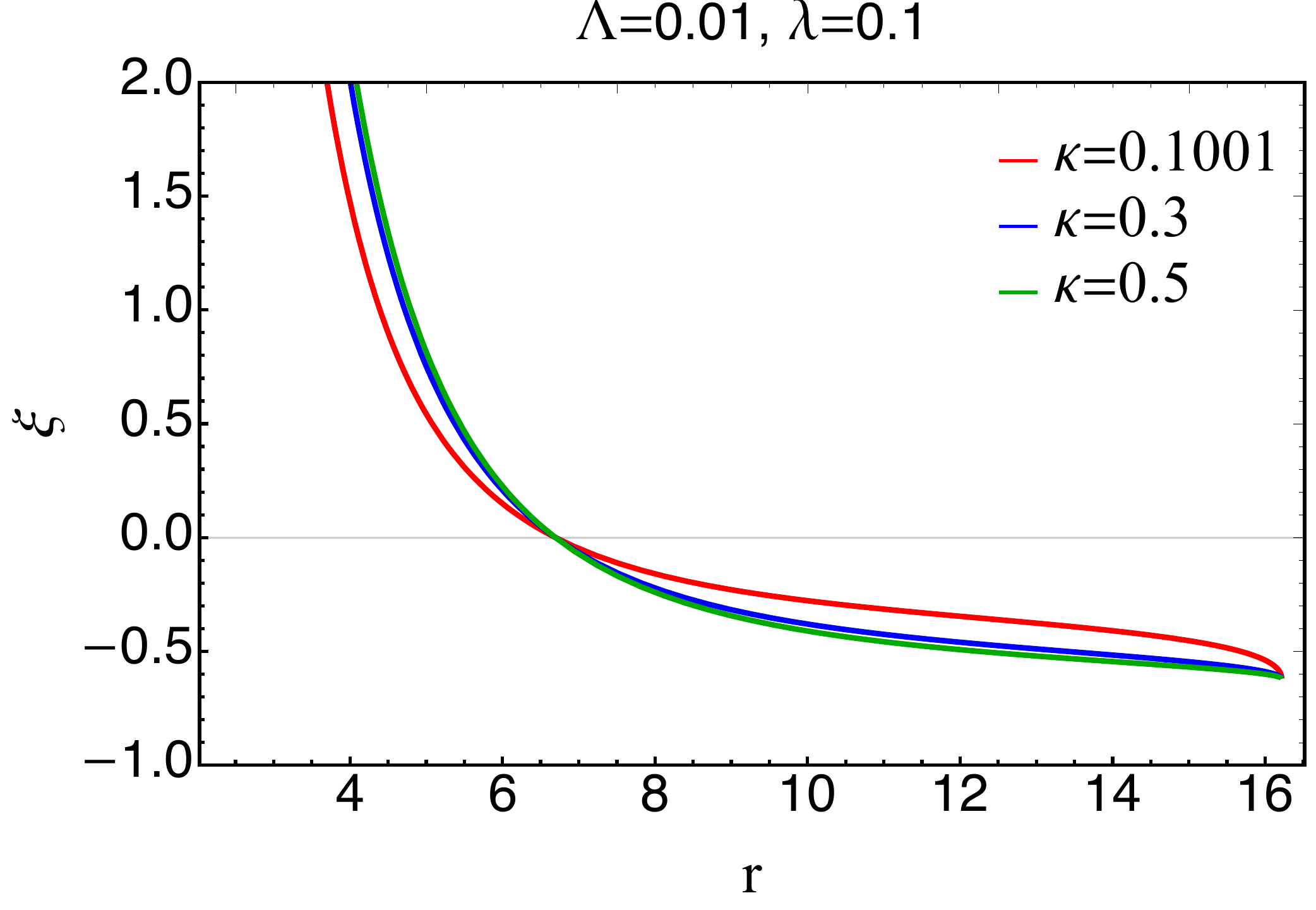}
	\endminipage
 \hfill
	%%%%%%%%%%%%%%%%%%%%%%%%
	\minipage{0.33\textwidth}
	\includegraphics[width=\linewidth]{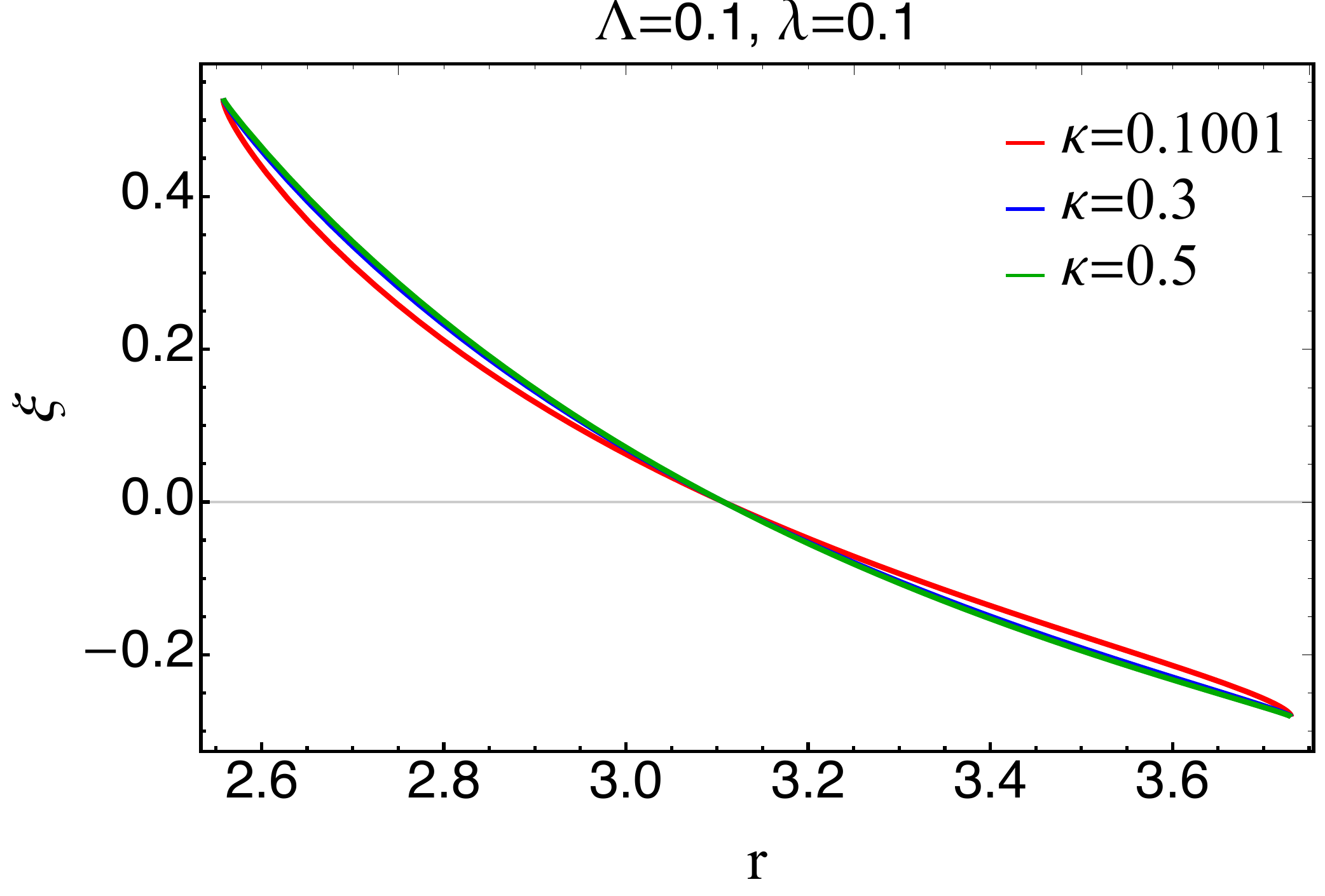}
	\endminipage
	\caption{The plot of exoticity parameter $\xi$ as function of $r$ in the range $r\in [\rth, r_c]$ for different values of $\Lambda$ and $\kappa$.}\label{fig_exoticity}
\end{figure*}	
%%%%%%%%%%%%%%%%%%%%%%%%%%%%%%%%%%%%%%%%%%%%%%%%%%%%%%%%%%%%%%%%%
%%%%%%%%%%%%%%%%%%%%%%%%%%%%%%%%%%%%%%%%%%%%%%%%%%%%%%%%%%%%%%%%%
%%%%%%%%%%%%%%%%%%%%%%%%%%%%%%%%%%%%%%%%%%%%%%%%%%%%%%%%%%%%
 \subsection{Violation of energy conditions}

The wormhole throat can alternatively defined by demanding that it is the location where the cross-sectional area of a bundle of radial null geodesics (i.e., the null congruence) is locally minimum \cite{Hochberg:1998ii}. This definition sheds light on the type of matter that supports a wormhole geometry.  To illustrate this, we can examine the Raychaudhuri equation for the null geodesics, given by \cite{Wald:1984rg, Poisson:2009pwt}
%%%%%
\begin{equation}
    \begin{aligned}
        \dfrac{d\hat\theta}{d\lambda}=-\frac{1}{2}\hat\theta^2-2\hat\sigma^2+2\hat\omega^2-R_{\mu\nu}k^{\mu}k^{\nu}~,
    \end{aligned}
\end{equation}
%%%%%
 $k^{\mu}$ represents the tangent to the null geodesics, and $\hat\theta$, $\hat\sigma$, and $\hat\omega$ denote the expansion (fractional change of cross-sectional area), shear, and vorticity of the bundle, respectively. In the case of a spherically symmetric spacetime with radially null congruence, the vorticity and shear of the bundle vanish identically. Furthermore, at the wormhole throat, we have $\hat\theta=0$ and $d\hat\theta/d\lambda\geq 0$ (the latter condition is known as the ``flare-out" condition \cite{Hochberg:1998ii}), as per the definition of the throat. This implies that $R_{\mu\nu}k^{\mu}k^{\nu}\leq 0$ at the throat. Using \autoref{einstein_equation}, we can deduce that this condition demands the effective energy-momentum tensor to violate the null energy condition near the throat. Thus, the construction of the wormholes requires the presence of \textit{exotic} matter that violates the null energy conditions \cite{Hochberg:1998ii}. Note that, in \cite{10.1119/1.15620}, Morris and Throne demanded that the exotic matter should violate weak energy conditions to have a wormhole solution. However, this distinction has minimal impact on our analysis, as the violation of the weak energy condition automatically implies the violation of the null energy condition \cite{Poisson:2009pwt}. Therefore, we shall proceed by examining whether the matter violates the weak energy condition or not.
 To investigate this, we introduce an orthonormal tetrad frame, denoted as $e^{\mu}_{(a)}$, where the latin indices refer to tetrad components. These components are raised or lowered using the metric $\eta^{ab}=(-1,1,1,1)$. Specifically, the tetrad frame can be expressed as follows
%%%%%%%%%%%%%%%%%%%%%%%%%%%%%%%%%%%%%%%%%%%%%%%%%%%%%%%%%%%%%%%%%
\begin{equation}\label{orthonormal_tetrad}
\begin{aligned}
e^{\mu}_{(0)}&=\left(\sqrt{-g^{00}},0,0,0\right)\,,\quad
e^{\mu}_{(1)}=\left(0,\sqrt{g^{11}},0,0\right)\\
e^{\mu}_{(2)}&=\left(0,0,\sqrt{g^{22}},0\right)\,,\quad
e^{\mu}_{(3)}=\left(0,0,0,\sqrt{g^{33}}\right)~.\\
\end{aligned}
\end{equation}
%%%%%%%%%%%%%%%%%%%%%%%%%%%%%%%%%%%%%%%%%%%%%%%%%%%%%%%%%%%%%%%%%
Using this orthonormal basis, we find the non-zero component of the effective energy-momentum tensor are
 %%%%%%%%%%%%%%%%%%%%%%%%%%%%%%%%%%%%%%%%%%%%%%%%%%%%%%%%%%%%%%%%%
\begin{equation}\label{SET_components}
\begin{aligned}
T_{(0)(0)}^{g}&=T_{\mu\nu}^g e^{\mu}_{(0)}e^{\mu}_{(0)}=\rho(r)\\
T_{(1)(1)}^{g}&=T_{\mu\nu}^g e^{\mu}_{(1)}e^{\mu}_{(1)}=-\tau(r)\\
T_{(2)(2)}^{g}&=T_{\mu\nu}^g e^{\mu}_{(2)}e^{\mu}_{(2)}=p(r)\\
T_{(3)(3)}^{g}&=T_{\mu\nu}^g e^{\mu}_{(3)}e^{\mu}_{(3)}=p(r)
\end{aligned}
\end{equation}
%%%%%%%%%%%%%%%%%%%%%%%%%%%%%%%%%%%%%%%%%%%%%%%%%%%%%%%%%%%%%%%%%
where $\rho(r)$ is the energy density, $\tau(r)$ is the radial tension and $p(r)$ is the tangential pressure. In this frame, the weak energy condition reduces to the following inequalities: $\rho\geq 0$, $\rho-\tau\geq 0$, and $\rho+p\geq 0$ \cite{Lemos:2003jb, Poisson:2009pwt}. For \autoref{bi-gravity_wormhole}, these conditions turns out to be 
%%%%%%%%%%%%%%%%%%%%%%%%%%%%%%%%%%%%%%%%%%%%%%%%%%%%%%%%%%%%%%%%%
\begin{equation}\label{energy_condition}
\begin{aligned}
\rho(r)&=\Lambda\\
\rho-\tau(r)&=\frac{2\kappa\left(\Lambda r^3-3M\right)}{3r^3\sqrt{-g_{00}}}\\
\rho+p(r)&=\frac{\kappa\left(2\Lambda r^3+3M\right)}{3r^3\sqrt{-g_{00}}}
\end{aligned}
\end{equation}
%%%%%%%%%%%%%%%%%%%%%%%%%%%%%%%%%%%%%%%%%%%%%%%%%%%%%%%%%%%%%%%%%
Clearly, for $\kappa> 0$ and $\Lambda>0$, $\rho> 0$, and $\rho+p> 0$ outside the wormhole throat. However, the condition $\rho-\tau\geq 0$ requires a little more attention. Following \cite{10.1119/1.15620}, we introduce the exoticity parameter, defined as $\xi=(\tau-\rho)/|\rho|$. The wormhole solution requires the imposition of the flare-out condition, i.e., $\xi>0$ at or in close proximity to the wormhole throat. In other words, the weak energy condition needs to be violated at or near the wormhole throat to obtain a valid wormhole solution. In \autoref{fig_exoticity}, we depict the behavior of $\xi$ as a function of $r$ in the range $r\in[\rth,r_c]$ for various values of $\Lambda$ and $\kappa$. The plot clearly illustrates that the weak energy condition is violated in the vicinity of the wormhole throat $\rth$. Interestingly, although the embedding surface becomes vertical, i.e., $dz/dr\to\infty$ (see \autoref{embedding}) at $r_c$, the weak energy condition is respected there. Thus, it rules out the possibility of having a second throat at $r_c$.\par

It is worth noting that a similar type of wormhole solution (but with $\Lambda=0$) was previously reported in the context of general relativity in \cite{Dadhich:2001fu}. In that work, the existence of such wormholes was established by imposing a specific condition on the energy-momentum tensor, namely $\rho=\rho_t=0$, where $\rho$ represents the density measured by a static observer and $\rho_t$ represents the convergence density experienced by a timelike congruence. Subsequently, it was subsequently found that such wormholes can also arise in brane-world scenarios \cite{Kar:2015lma}. It is interesting to note that such kind of solution can be obtained in the bi-gravity theory discussed in this paper with $\beta_{1} = \beta_{4}$. 
	%%%%%%%%%%%%%%%%%%%%%%%%%%%%%%%%%%%%%%%%%%%%%%%%%%%%%%%%%%%%%%%%%%%%%%%%%%%%%%%%%%%%%%%%%%%%%%%%%%%%%
% background. %%%%%%%%%%%%%%%%%%%%%%%%%%%%%%%%%%%%%%%%%%%%%%%%%%%%%%%%%%%%%%%%%%%%%%%%%%%%%%%%%%%%%%%%%%%%%%%%%%%
\subsection{Traversability Criteria}
%%%%%%%%%%%%%%%%%%%%%%%%%%%%%%%%%%%%%%%%%%%%%%%%%%%%%%%%%%%%%%%%%%%%%%%%%%%%%%%%%%%%%%%%%%%%%%%%%%%
In this section, we discuss whether the proposed wormhole allows safe interstellar travel for a human being. To ensure safe passage through the wormhole, Morris and Thorne put forward the following conditions \cite{10.1119/1.15620}: Firstly, the acceleration experienced by the traveler should be comparable to the gravitational acceleration on Earth surface $g_{\oplus}(=9.8~\textrm{m s}^{-1}$ in SI unit). Secondly, the tidal acceleration exerted on different parts of the traveler's body should also be in the same order as $g_{\oplus}$. For the sake of simplicity, we consider a traveler moving radially with four-velocity $v^{\mu}$. The first criterion for such a traveler can be expressed as follows \cite{10.1119/1.15620}
 %%%%%%%%%%%%%%%%%%%%%%%%%%%%%%%%%%%%%%%%%%%%%%%%%%%%%%%%%%%%%%%%%%%%%%%%%%%%%%%%%%%%%%%%%%%%%%%%%%%
\begin{equation}\label{acce1}
    \left|\sqrt{\frac{G(r)}{F(r)}} \frac{d \left(\gamma  \sqrt{F(r)}\right)}{d r}\right|\lesssim g_{\oplus}\approx \frac{1}{1~\textrm{light year}}
\end{equation}
%background. %%%%%%%%%%%%%%%%%%%%%%%%%%%%%%%%%%%%%%%%%%%%%%%%%%%%%%%%%%%%%%%%%%%%%%%%%%%%%%%%%%%%%%%%%%%%%%%%%%%
where, $F(r)=-g_{00}$, $G(r)=1/g_{11}$ and $\gamma=1/\sqrt{1-v^2}$. If we consider a traveler of height $2~\textrm{m}$, the second criterion turns out to be \cite{10.1119/1.15620, Visser:1995cc}
 %%%%%%%%%%%%%%%%%%%%%%%%%%%%%%%%%%%%%%%%%%%%%%%%%%%%%%%%%%%%%%%%%%%%%%%%%%%%%%%%%%%%%%%%%%%%%%%%%%%
 \begin{widetext}
\begin{equation}\label{acce2}
\begin{aligned}
\left|R_{(0)(1)(0)(1)}\right|&=
   \bigg|\frac{F'(r) G'(r)}{4 F(r)}-\frac{G(r) F'(r)^2}{4 F(r)^2}+\frac{G(r) F''(r)}{2 F(r)}\bigg|\lesssim  \frac{g_{\oplus}}{2~\textrm{m}}\approx \frac{1}{(10^5~\textrm{km})^2} \\  
   \left|R_{(0)(2)(0)(2)}\right|&=
   \bigg|\gamma ^2 \left(\frac{G(r) F'(r)}{2 r F(r)}-\frac{v^2 G'(r)}{2 r}\right)\bigg|\lesssim  \frac{g_{\oplus}}{2~\textrm{m}}\approx \frac{1}{(10^5~\textrm{km})^2}
\end{aligned}
\end{equation}
\end{widetext}
%background. %%%%%%%%%%%%%%%%%%%%%%%%%%%%%%%%%%%%%%%%%%%%%%%%%%%%%%%%%%%%%%%%%%%%%%%%%%%%%%%%%%%%%%%%%%%%%%%%%%%
where the Riemann tensor components are calculated in the orthonormal tetrad frame (see \autoref{orthonormal_tetrad}). At the wormhole throat, \autoref{acce1} and \autoref{acce2} takes a particularly simpler form
\begin{equation}\label{acce3}
\begin{aligned}
   \bigg|\frac{\gamma  \lambda  \left(\Lambda  \rth^3-3 M\right)}{3 \kappa  \rth^2}\bigg|&\lesssim  \frac{1}{1~\textrm{light year}} \\
   \bigg|\frac{1}{3} v^2 \gamma ^2 \left(\frac{3 M}{\rth^3}-\Lambda \right)\bigg|&\lesssim  \frac{1}{(10^5~\textrm{km})^2}
\end{aligned}
\end{equation}
%%%%%%%%%%%%%%%%%%%%%%%%%%%%%%%%%%%%%%%%%%%%%%%%%%%%%%%%%%%%%%%%%%%%%%%%%%%%%%%%%%%%%%%%%%%%%%%%%%%
Noting that the measured value of the cosmological constant is approximately $10^{-52}~\textrm{m}^{-2}$\cite{PhysRevD.86.010001}, we can ignore the contribution of the $\Lambda$ in the above equation. In that scenario, the wormhole throat is located at $\rth\approx 2M$. If we take $\rth\approx 2M=10^5~\textrm{km}$, we find that $\kappa/\lambda\gtrsim 10^8 \gamma$ and $\beta^2\gamma^2\lesssim 2$ which implies $v\lesssim \sqrt{2/3}$. Therefore, the first criterion in \autoref{acce3} determines the allowed parameter range of $\kappa$ and $\lambda$ for a traversable wormhole of a given mass, while the second criterion constrains the velocity at which the traveler can cross the wormhole.
%background. %%%%%%%%%%%%%%%%%%%%%%%%%%%%%%%%%%%%%%%%%%%%%%%%%%%%%%%%%%%%%%%%%%%%%%%%%%%%%%%%%%%%%%%%%%%%%%%%%%%
%%%%%%%%%%%%%%%%%%%%%%%%%%%%%%%%%%%%%%%%%%%%%%%%%%%%%%%%%%%%%%%%%%%%%%%%%%%%%%%%%%%%%%%%%%%%%%%%%%%
 \section{Linear perturbations of wormholes in bi-gravity theories}\label{Sec:Perurbation}

In this section, we study the perturbation of the wormhole spacetime by scalar and electromagnetic fields.
Here, we restrict our attention to linear perturbation approximation, which ensures that the perturbing field does not backreact on the background spacetime.     
%In this section, we discuss the propagation of scalar and electromagnetic field in the \BGBH\ background. %%%%%%%%%%%%%%%%%%%%%%%%%%%%%%%%%%%%%%%%%%%%%%%%%%%%%%%%%%%%%%%%%%%%%%%%%%%%%%%%%%%%%%%%%%%%%%%%%%%
%%%%%%%%%%%%%%%%%%%%%%%%%%%%%%%%%%%%%%%%%%%%%%%%%%%%%%%%%%%%%%%%%%%%%%%%%%%%%%%%%%%%%%%%%%%%%%%%%%%
\begin{figure*}[htb!]
	%%%%%%%%%%%%%%%%%%%%%%%%
	\centering
	\minipage{0.48\textwidth}
	\includegraphics[width=\linewidth]{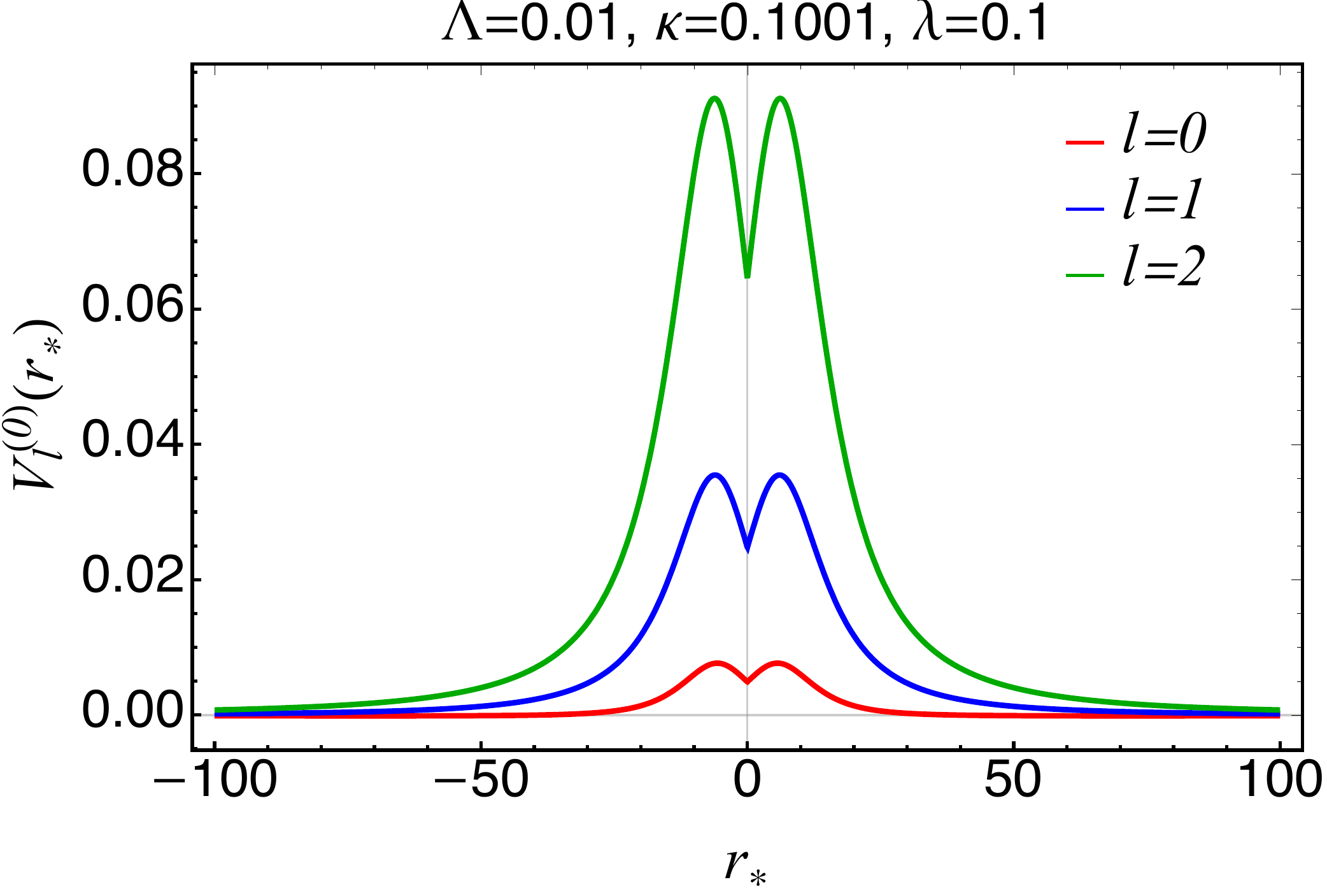}
	\endminipage\hfill
	%%%%%%%%%%%%%%%%%%%%%%%%
	\minipage{0.48\textwidth}
	\includegraphics[width=\linewidth]{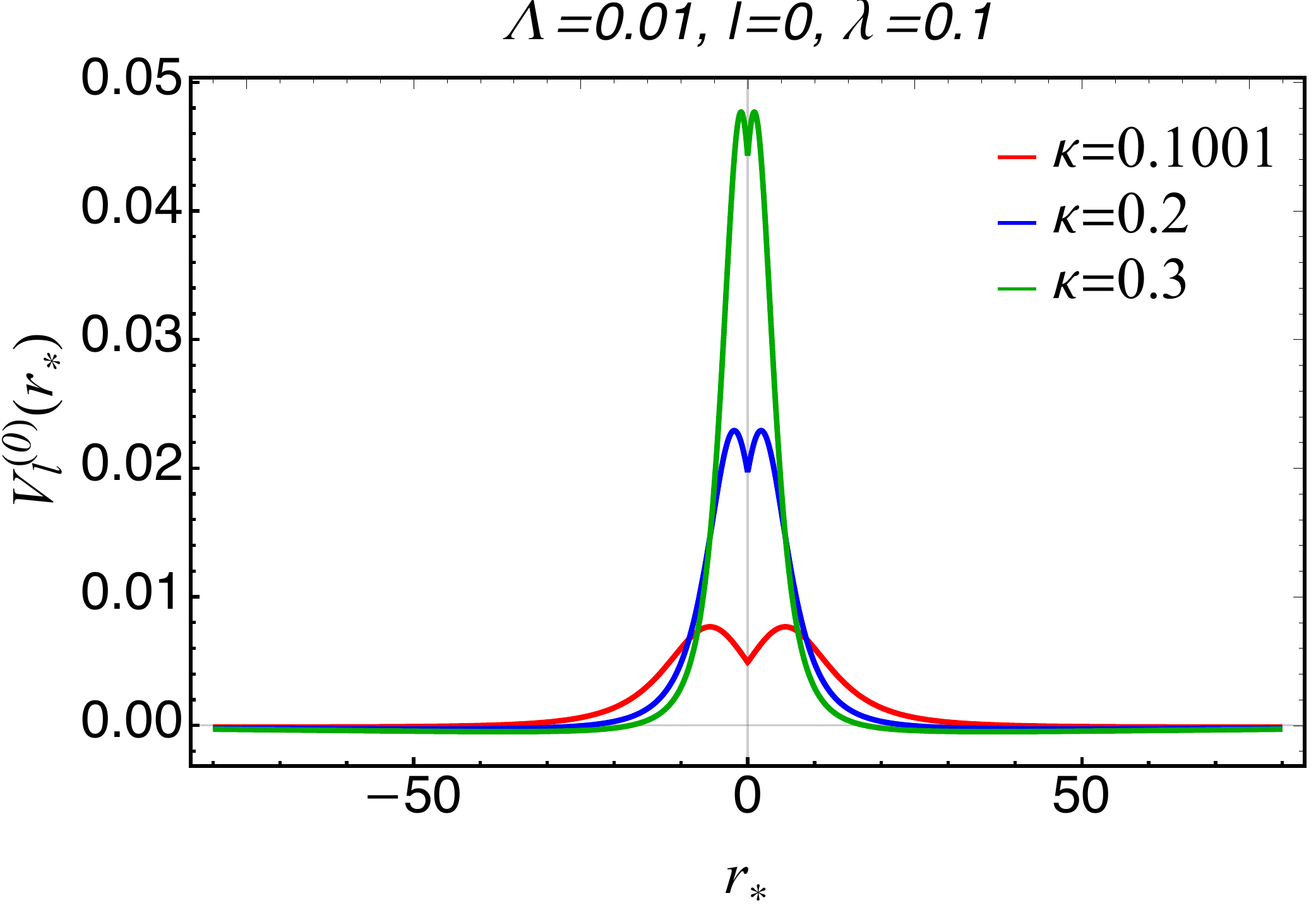}
	\endminipage
	\caption{The plot of scalar field perturbation potential $V_{l}^{(0)}$ as a function of $r_{*}$ for $\Lambda=0.01$, $\lambda=0.1$ and different values of $l$ (left panel ) and $\kappa$ (right panel). }\label{fig_scalar_pot}
\end{figure*}	
%%%%%%%%%%%%%%%%%%%%%%%%%%%%%%%%%%%%%%%%%%%%%%%%%%%%%%%%%%%%%%%%%
%%%%%%%%%%%%%%%%%%%%%%%%%%%%%%%%%%%%%%%%%%%%%%%%%%%%%%%%%%%%%%%%%%%%%%%%%%%%%%%%%%%%%%%%%%%%%%%%%%%
\begin{figure*}[htb!]
	%%%%%%%%%%%%%%%%%%%%%%%%
	\centering
	\minipage{0.48\textwidth}
	\includegraphics[width=\linewidth]{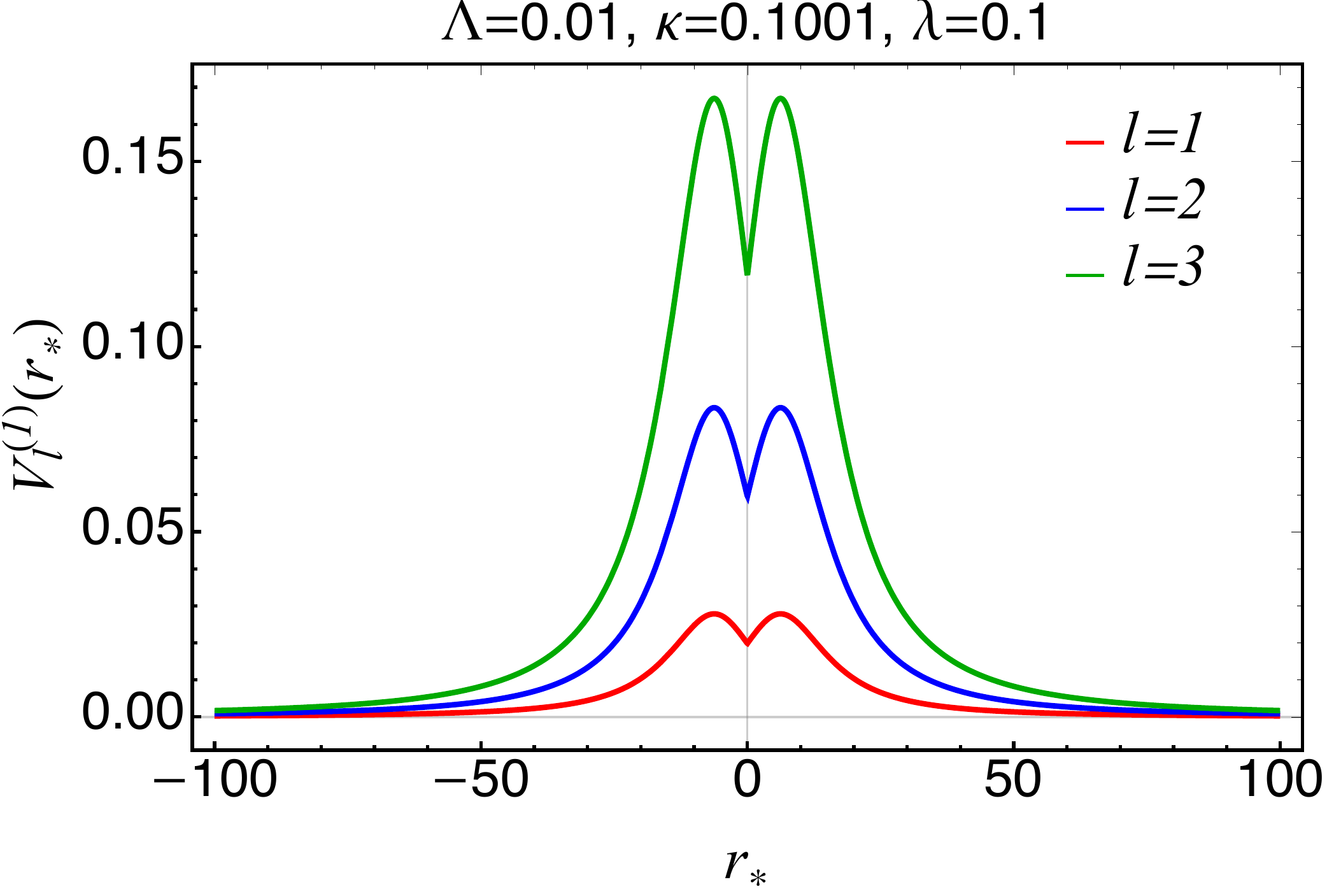}
	\endminipage\hfill
	%%%%%%%%%%%%%%%%%%%%%%%%
	\minipage{0.48\textwidth}
	\includegraphics[width=\linewidth]{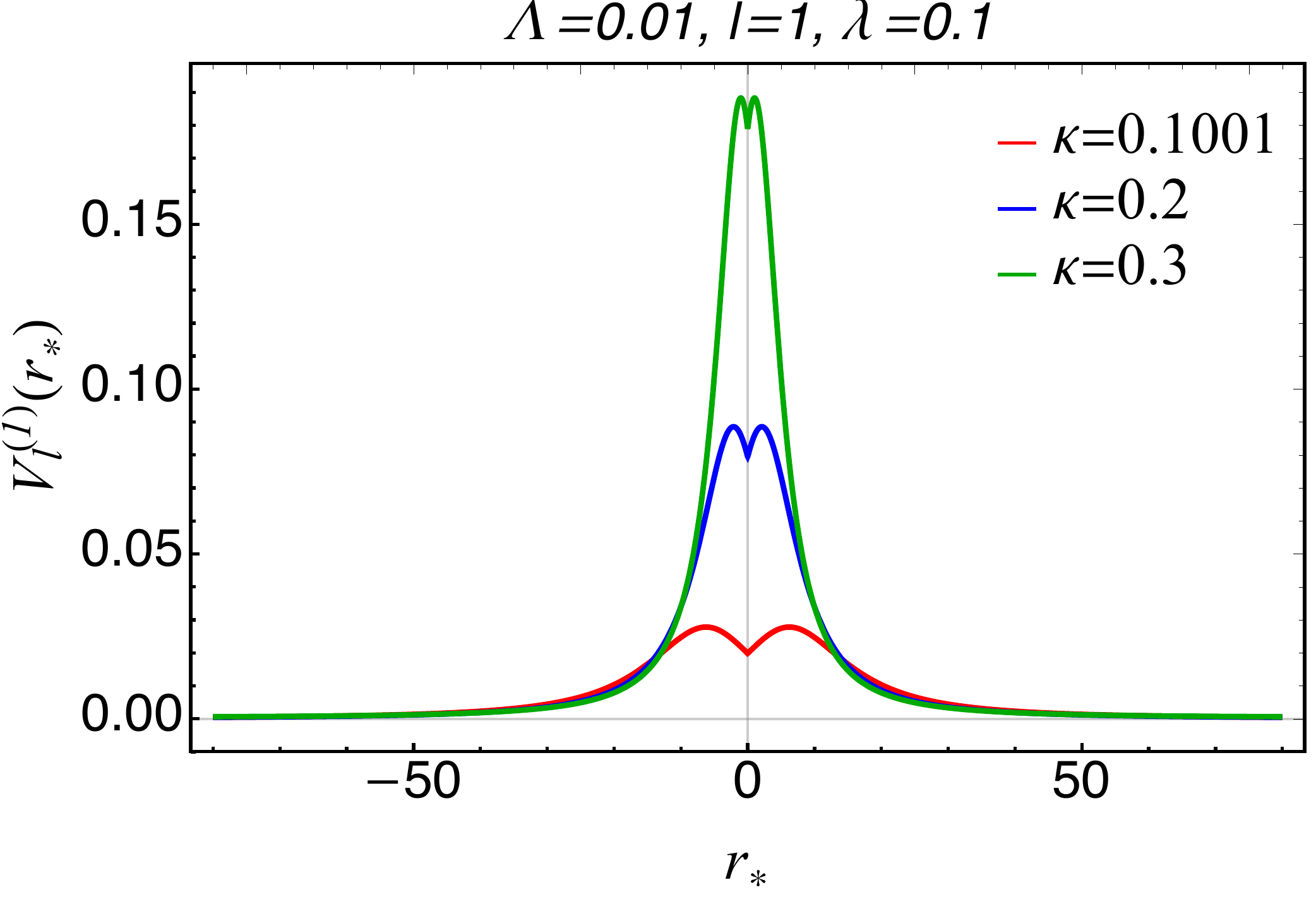}
	\endminipage
	\caption{The plot of electromagnetic field perturbation potential $V_{l}^{(1)}$ as a function of $r_{*}$ for $\Lambda=0.01$, $\lambda=0.1$ and different values of $l$ (left panel ) and $\kappa$ (right panel).}\label{fig_em_pot}
\end{figure*}	
%%%%%%%%%%%%%%%%%%%%%%%%%%%%%%%%%%%%%%%%%%%%%%%%%%%%%%%%%%%%%%%%%
\subsection{Massless, minimally coupled Scalar field}

Under the linear perturbation approximation scheme the perturbation equation of the field reduces to the equation of motion of the corresponding field in the background spacetime. For a  massless, minimally coupled scalar field, this is governed by the Klein-Gordon equation 
%%%%%%%%%%%%%%%%%%%%%%%%%%%%%%%%%%
\begin{equation}
\begin{aligned}\label{KG_Eqn}
\Box \Phi=\frac{1}{\sqrt{-g}}\partial_\mu\left(\sqrt{-g}g^{\mu\nu}\partial_\nu \Phi\right)=0~.
\end{aligned}
\end{equation}
%%%%%%%%%%%%%%%%%%%%%%%%%%%%%%%%%%
The static and spherically symmetric nature of the background spacetime allows us decompose the scalar field in the following manner 
%%%%%%%%%%%%%%%%%%%%%%%%%%%%%%%%%%
\begin{equation}
\begin{aligned}\label{Scalar_decomposition}
\Phi(t,r,\theta,\phi)=\sum_{l=0}^{\infty}\sum_{m=-l}^{l}\frac{\psi_{\ell m}^{(0)}(r)}{r}e^{-i \omega t}Y_{lm}(\theta,\phi)~,
\end{aligned}
\end{equation}
%%%%%%%%%%%%%%%%%%%%%%%%%%%%%%%%%%
where, $Y_{lm}(\theta,\phi)$ is the spherical harmonics and $\psi_{\ell m}^{(0)}$ is the radial master function. Replacing \autoref{Scalar_decomposition} in \autoref{KG_Eqn}, we can cast the radial perturbation equation in the form of Schr\"{o}dinger equation
%%%%%%%%%%%%%%%%%%%%%%%%%%%%%%%%%%
\begin{align}\label{Radial_scalar}
\frac{d^{2}\psi_{lm}^{(0)}}{dr_{*}^{2}}+[\omega^{2}-V_{l}^{(0)}(r)]\psi_{lm}^{(0)}=0~,
\end{align}  
%%%%%%%%%%%%%%%%%%%%%%%%%%%%%%%%%%
where,  $V_{l}^{(0)}(r)$ is the radial perturbation potential which can be written as follows \cite{Biswas:2022wah},
\begin{equation}\label{Scalar_potential}
   V_{l}^{(0)}(r)= e^{2a} \frac{l(l+1) }{r^2}+\frac{ e^{(a-b)}}{r}\frac{d \left(e^{(a-b)}\right)}{dr}~.
\end{equation}
%%%%%%%%%%%%%%%%%%%%%%%%%%%%%%%%%%%%%%%%%%%%%%%%%%%%%%%%%%%%%%%%%%%%%%%%%%%%%%%%%%%%%%%%%%%%%%%%%%%
In \autoref{Radial_scalar}, the symbol $r_{*}$ represents the tortoise coordinate, which can be obtained by solving the differential equation
%%%%%%%%%%%%%%%%%%%%%%%%%%%%%%%%%%
\begin{equation}\label{tortoise}
\begin{aligned}
\frac{dr_{*}}{dr}=\exp\left[b(r)-a(r)\right]~,
\end{aligned}
\end{equation}
%%%%%%%%%%%%%%%%%%%%%%%%%%%%%%%%%%
where the boundary condition is chosen such that the tortoise coordinate vanishes at the wormhole throat, i.e., $r_{*}=0$ at $r=\rth$. The definition of the tortoise coordinate is useful because it enables us to describe the wormhole geometry as if two black hole spacetimes were joined together at the throat, with the tortoise coordinate covering both universes on either side of the throat. We solve the differential equation \autoref{tortoise} using \texttt{Mathematica} to obtain the tortoise coordinate \cite{Mathematica}.\\
%%%%%%%%%%%%%%%%%%%%%%%%%%%%%%%%%%
In \autoref{fig_scalar_pot}, we plot scalar field perturbation potential  $V_{l}^{(0)}$ as a function of tortoise coordinate $r_{*}$ for different values of angular number $l$ and $\kappa$. Note that the perturbation potential has a double-bump structure that is symmetric about the wormhole throat ($r_{*}=0$). 
%%%%%%%%%%%%%%%%%%%%%%%%%%%%%%%%%%%%%%%%%%%%%%%%%%%%%%%%%%%%%%%%%%%%%%%%%%%%%%%%%%%%%%%%%%%%%%%%%%%
%%%%%%%%%%%%%%%%%%%%%%%%%%%%%%%%%%%%%%%%%%%%%%%%%%%%%%%%%%%%%%%%%%%%%%%%
%%%%%%%%%%%%%%%%%%%%%%%%%%%%%%%%%%%%%%%%%%%%%%%%%%%%%%%%%%%%%%%%%%%%%%%%%%%%%%%%%%%%%%%%%%%%%%%%%%%
\subsection{Electromagnetic field}

The evolution of the massless vector field perturbation is governed by the Maxwell's equation
%%%%%%%%%%%%%%%%%%%%%%%%%%%%%%%%%%
\begin{equation}
\begin{aligned}\label{Maxwell_Eqn}
\nabla_{\nu}F^{\mu\nu}=0~,
\end{aligned}
\end{equation}
%%%%%%%%%%%%%%%%%%%%%%%%%%%%%%%%%%
where, $\nabla_\mu$ denotes covarient derivative with respect to the background metric and $F_{\mu\nu}=\partial_\mu \mA_\nu-\partial_\nu \mA_\mu$ is the Maxwell field tensor. Here, $\mA^{\mu}$ is the vector potential. Owing to the spherical symmetry of the background spacetime, we can decompose the vector fields as\cite{PhysRevD.9.860}
%%%%%%%%%%%%%%%%%%%%%%%%%%%%%%%%%%
\begin{equation}\label{A-decomp}
 \begin{aligned}
\mA_{\mu}(t,r,\theta,\phi)=\int d\omega \sum_{l,m}\Bigg[&\alpha_{lm}(r)e^{-i\omega t}\begin{pmatrix}
0\\
0\\
\frac{1}{\sin\theta}\partial_{\phi}Y_{lm}\\
-\sin\theta\partial_{\theta}Y_{lm}
\end{pmatrix}_{\rm odd}\\&+
e^{-i\omega t}\begin{pmatrix}
f_{lm}(r)Y_{lm}\\
u_{lm}(r)Y_{lm}\\
k_{lm}(r)\partial_{\theta}Y_{lm}\\
k_{lm}(r)\partial_{\phi}Y_{lm}
\end{pmatrix}_{\rm even}\Bigg]~.
\end{aligned}   
\end{equation}
%%%%%%%%%%%%%%%%%%%%%%%%%%%%%%%%%%
Here, the first term in the right hand side has parity $(-1)^{l+1}$ (odd parity), whereas the second term has parity $(-1)^{l}$ (even parity). Inserting \autoref{A-decomp} into \autoref{Maxwell_Eqn}, we find that perturbation equation for odd and even parity sector can be written as follows \cite{Biswas:2022wah}
%%%%%%%%%%%%%%%%%%%%%%%%%%%%%%%%%%
\begin{align}\label{Radial_em}
\frac{d^{2}\psi_{lm}^{(1)}}{dr_{*}^{2}}+[\omega^{2}-V_{l}^{(1)}(r)]\psi_{lm}^{(1)}=0~,
\end{align}  
%%%%%%%%%%%%%%%%%%%%%%%%%%%%%%%%%%
where  $V_{l}^{(1)}(r)$ is the radial perturbation potential which can be written as follows,
\begin{equation}\label{em_potential}
   V_{l}^{(1)}(r)= e^{2a} \left[\frac{l(l+1) }{r^2}\right]~.
\end{equation}
%%%%%%%%%%%%%%%%%%%%%%%%%%%%%%%%%%%%%%%%%%%%%%%%%%%%%%%%%%%%%%%%%%%%%%%%%%%%%%%%%%%%%%%%%%%%%%%%%%%
Here, the master functions in the odd and even parity sector are given by the following expression
\begin{equation}
\psi_{lm}^{(1)}=\begin{cases} \alpha_{lm}, & \textrm{odd parity}\\
\frac{r^2}{l(l+1)}(-i\omega u_{lm}-\frac{df_{lm}}{dr})& \textrm{even parity}\end{cases}
\end{equation}
%%%%%%%%%%%%%%%%%%%%%%%%%%%%%%%%%%%%%%%%%%%%%%%%%%%%%%%%%%%%%%%%%%%%%%%%%%%%%%%%%%%%%%%%%%%%%%%%%%%
In \autoref{fig_em_pot}, we plot electromagnetic field perturbation potential  $V_{l}^{(1)}$ as a function of tortoise coordinate $r_{*}$ for different values of angular number $l$ and $\kappa$. Similar to the scalar field case, the electromagnetic field perturbation potential has a double-bump structure that is symmetric about the wormhole throat ($r_{*}=0$). 
%%%%%%%%%%%%%%%%%%%%%%%%%%%%%%%%%%%%%%%%%%%%%%%%%%%%%%%%%%%%%%%%%%%%%%%%%%%%%%%%%%%%%%%%%%%%%%%%%%%
%%%%%%%%%%%%%%%%%%%%%%%%%%%%%%%%%%%%%%%%%%%%%%%%%%%%%%%%%%%%%%%%%%%%%%%%%%%%%%%%%%%%%%%%%%%%%%%%%%%

	%\includegraphics[width=\linewidth]{latetime_k1001_l0.pdf}
	%\includegraphics[width=\linewidth]{latetime_k1001_l4.pdf}
%%%%%%%%%%%%%%%%%%%%%%%%%%%%%%%%%%%%%%%%%%%%%%%%%%%%%%%%%%%%%%%%%
%%%%%%%%%%%%%%%%%%%%%%%%%%%%%%%%%%%%%%%%%%%%%%%%%%%%%%%%%%%%%%%%%
%%%%%%%%%%%%%%%%%%%%%%%%%%%%%%%%%%%%%%%%%%%%%%%%%%%%%%%%%%%%%%%%%
\begin{figure*}[htb!]
	%%%%%%%%%%%%%%%%%%%%%%%%
	\centering
	\minipage{0.33\textwidth}
	\includegraphics[width=\linewidth]{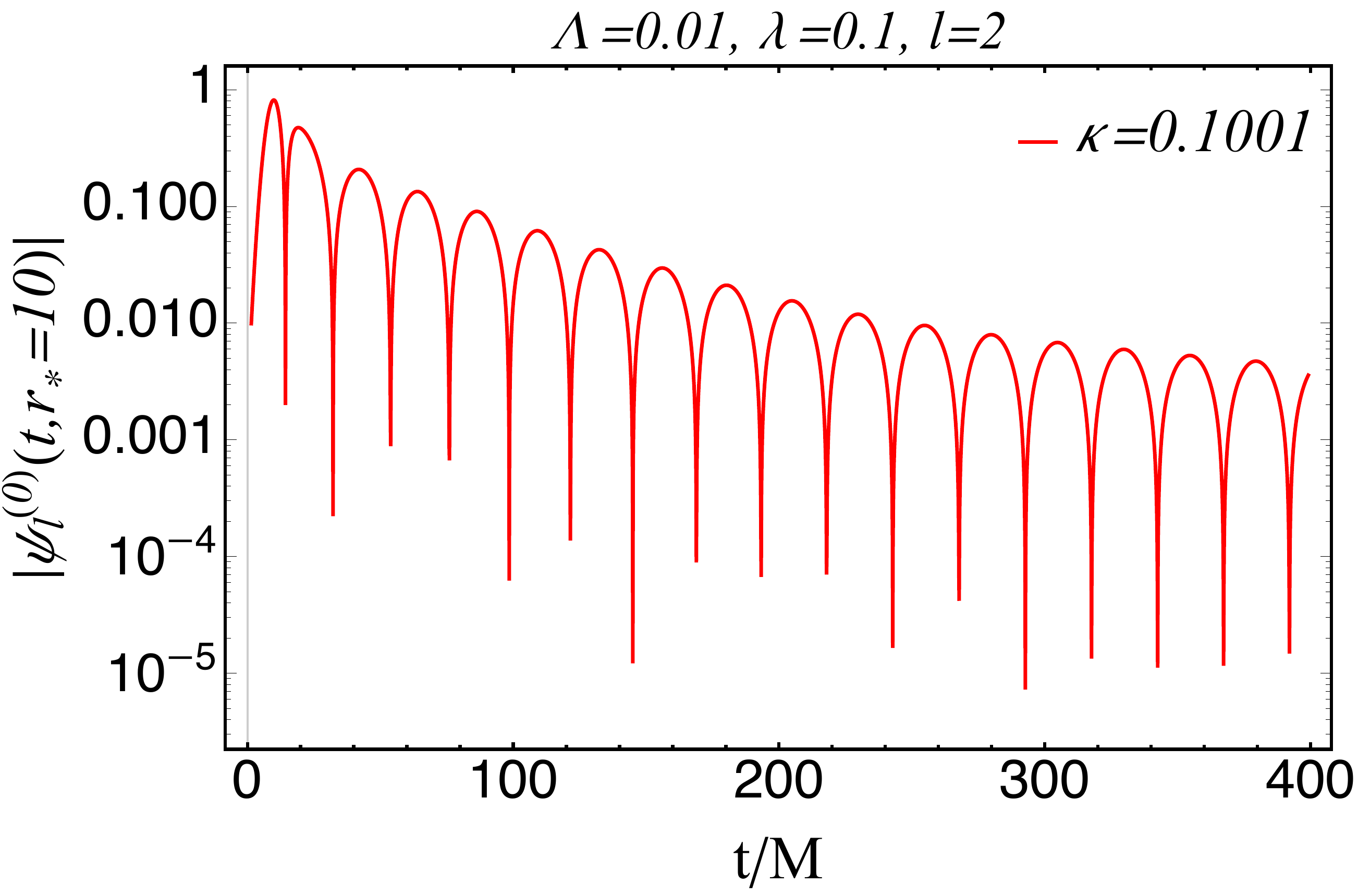}
	\endminipage\hfill
	%%%%%%%%%%%%%%%%%%%%%%%%
	\minipage{0.33\textwidth}
	\includegraphics[width=\linewidth]{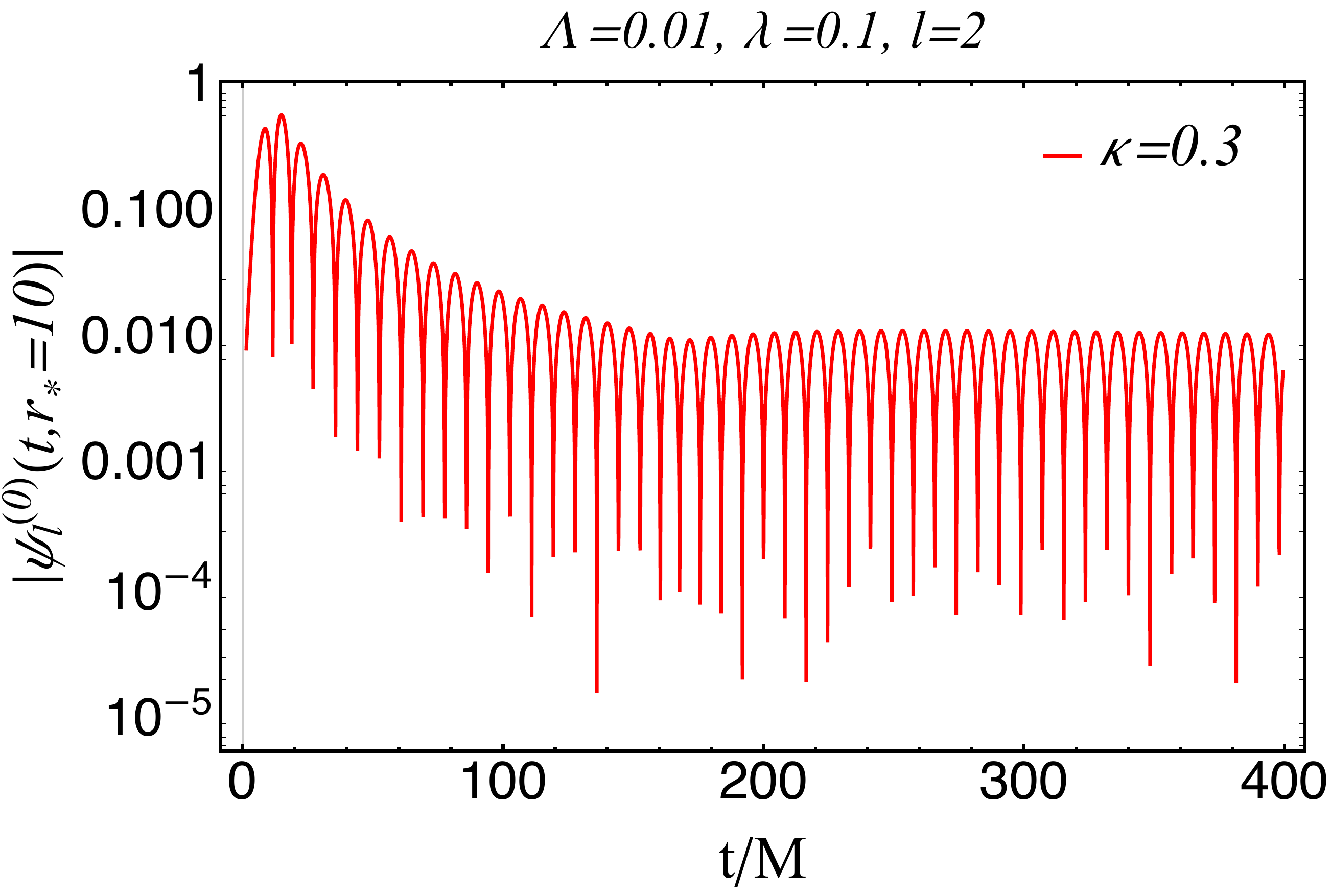}
	\endminipage
 \hfill
	%%%%%%%%%%%%%%%%%%%%%%%%
	\minipage{0.33\textwidth}
	\includegraphics[width=\linewidth]{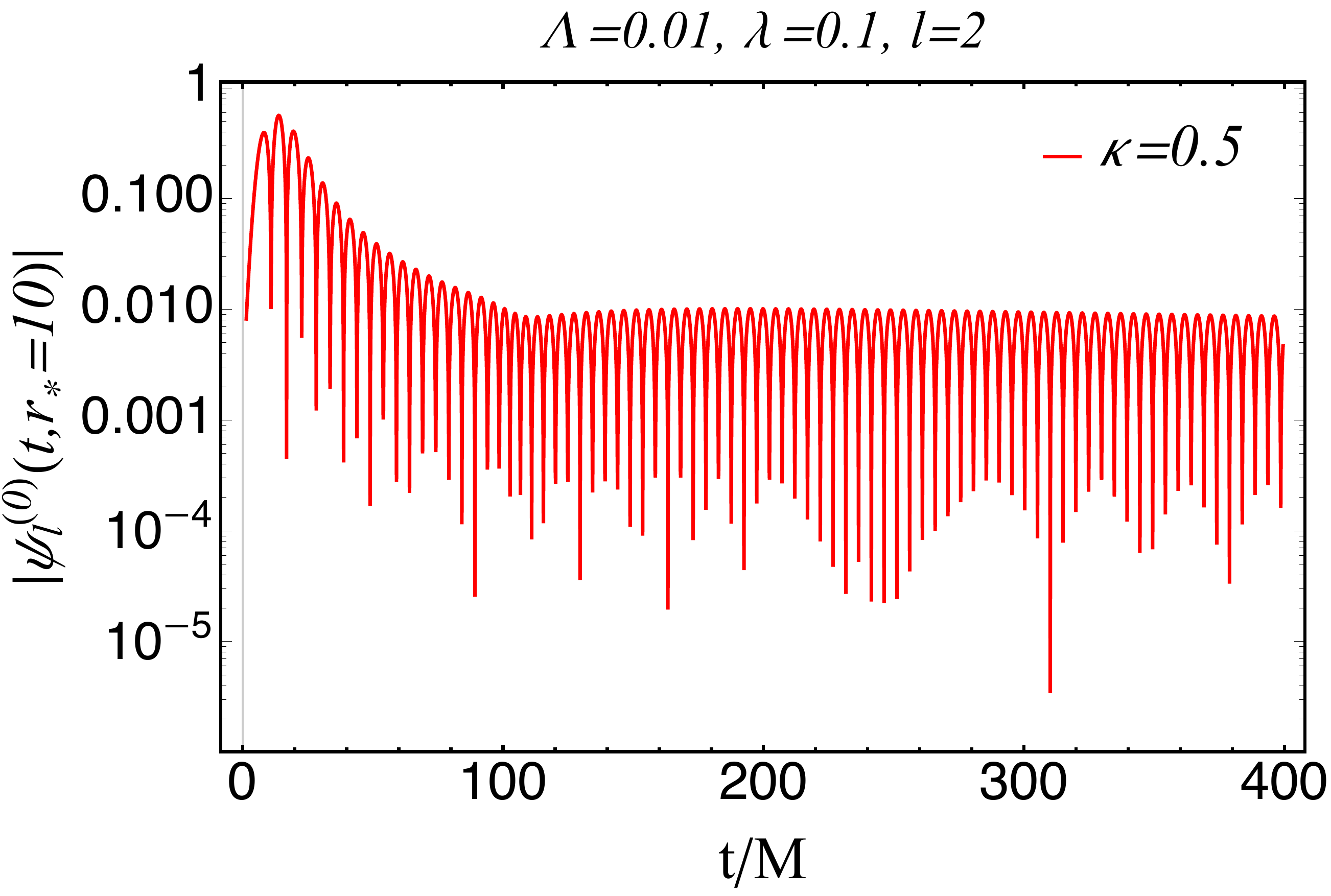}
	\endminipage\hfill
 \minipage{0.33\textwidth}
	\includegraphics[width=\linewidth]{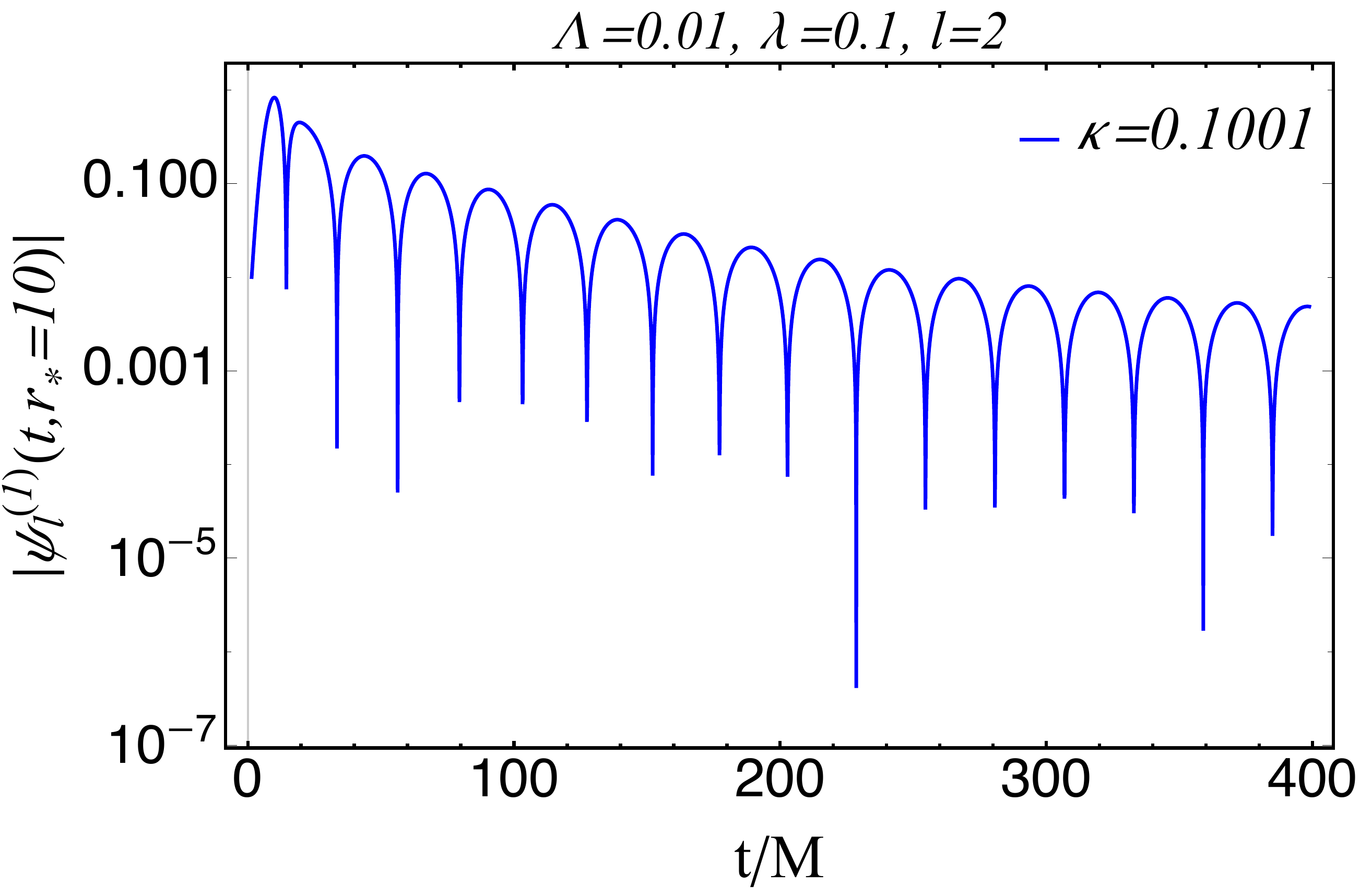}
	\endminipage\hfill
	%%%%%%%%%%%%%%%%%%%%%%%%
	\minipage{0.33\textwidth}
	\includegraphics[width=\linewidth]{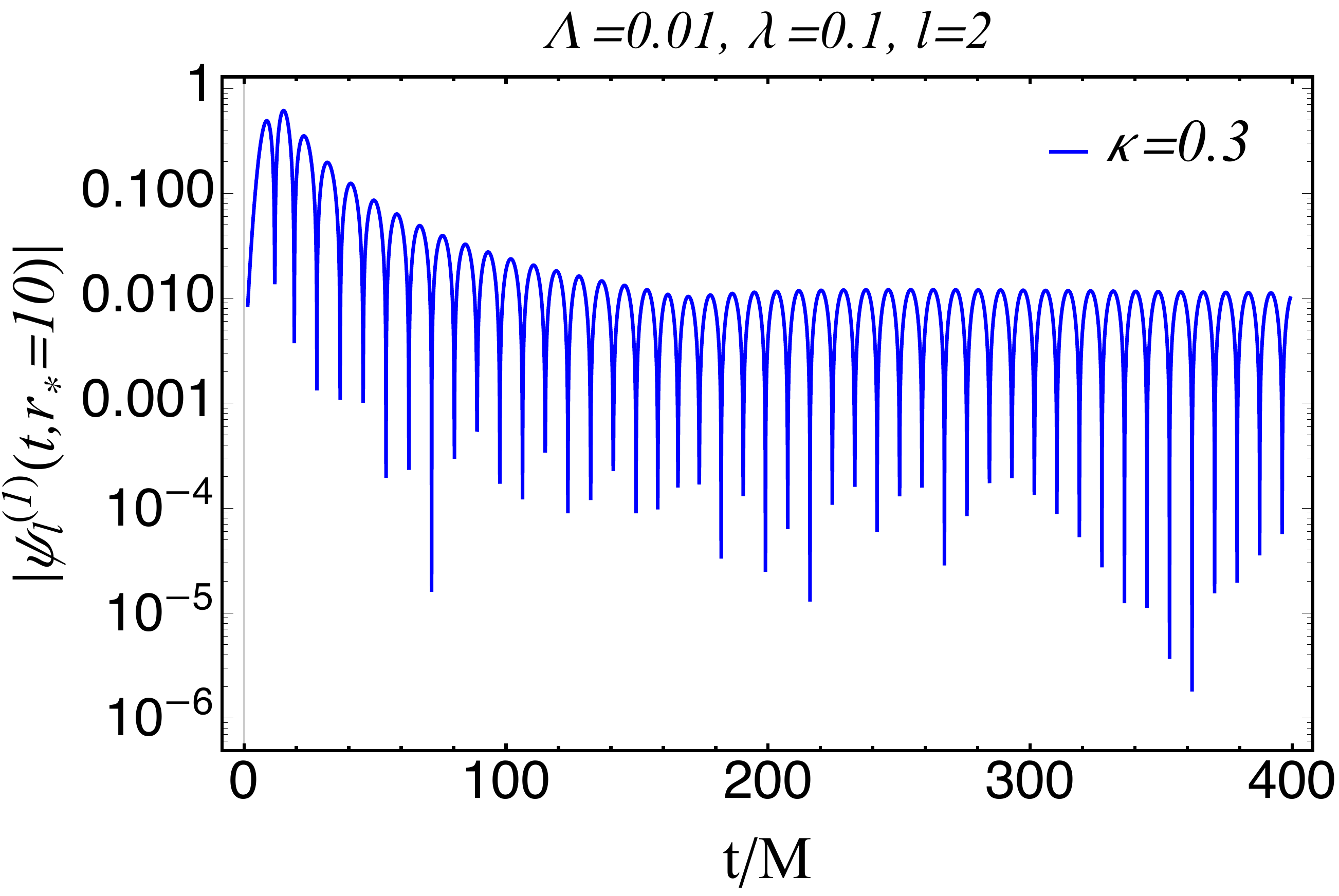}
	\endminipage
 \hfill
	%%%%%%%%%%%%%%%%%%%%%%%%
	\minipage{0.33\textwidth}
	\includegraphics[width=\linewidth]{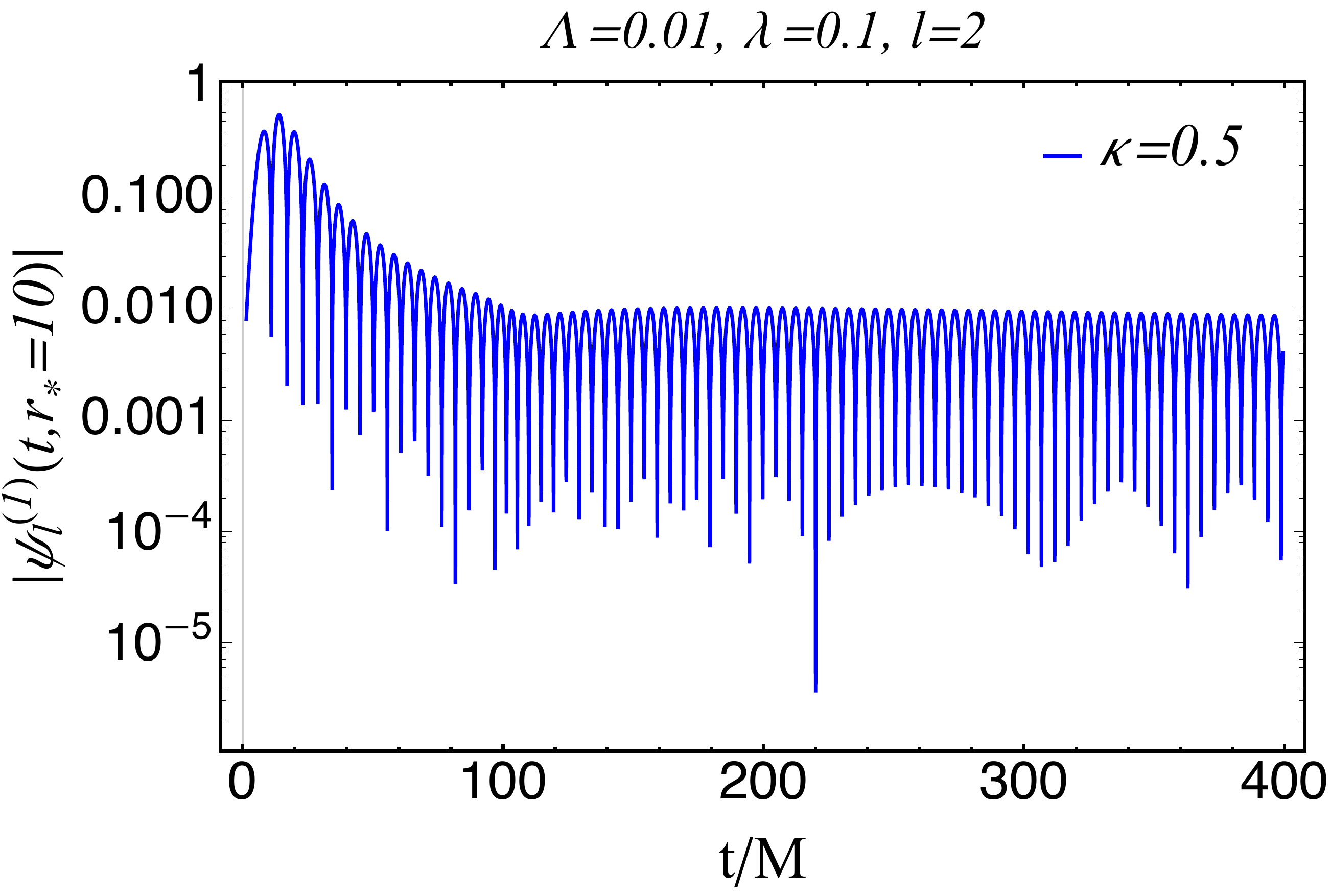}
	\endminipage
	\caption{Upper panel: Time evolution of scalar perturbation in the wormhole background for different values of $\kappa$. Lower panel: The same for electromagnetic perturbation. In both of these cases, we consider $M=1$, $\Lambda=0.01$, $l=2$, and $\lambda=0.1$.}\label{fig_latetime_different_kappa}
\end{figure*}	
%%%%%%%%%%%%%%%%%%%%%%%%%%%%%%%%%%%%%%%%%%%%%%%%%%%%%%%%%%%%%%%%%
%%%%%%%%%%%%%%%%%%%%%%%%%%%%%%%%%%%%%%%%%%%%%%%%%%%%%%%%%%%%%%%%%
%%%%%%%%%%%%%%%%%%%%%%%%%%%%%%%%%%%%%%%%%%%%%%%%%%%%%%%%%%%%%%%%%
\begin{figure*}[htb!]
	%%%%%%%%%%%%%%%%%%%%%%%%
	\centering
	\minipage{0.33\textwidth}
	\includegraphics[width=\linewidth]{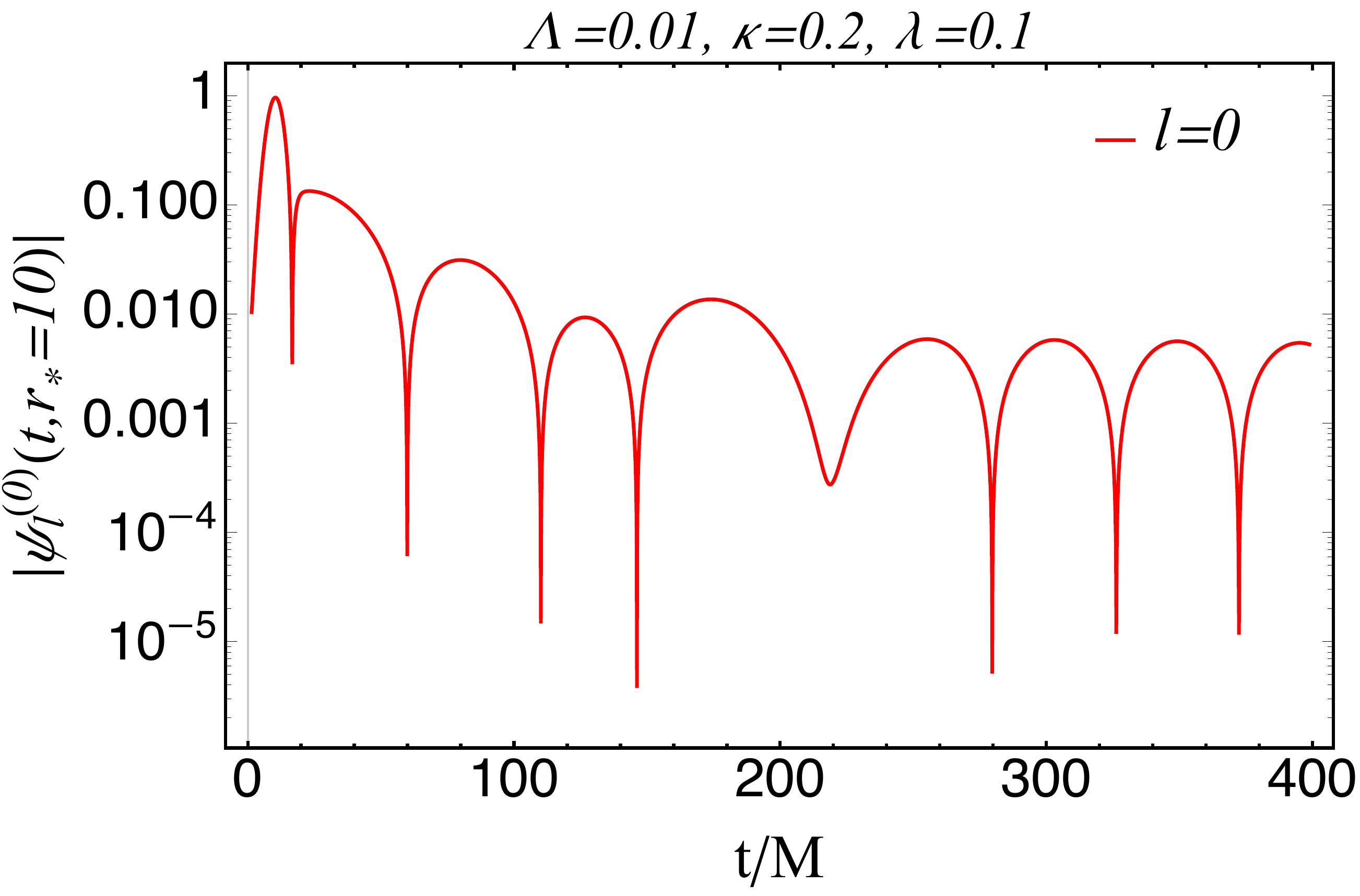}
	\endminipage\hfill
	%%%%%%%%%%%%%%%%%%%%%%%%
	\minipage{0.33\textwidth}
	\includegraphics[width=\linewidth]{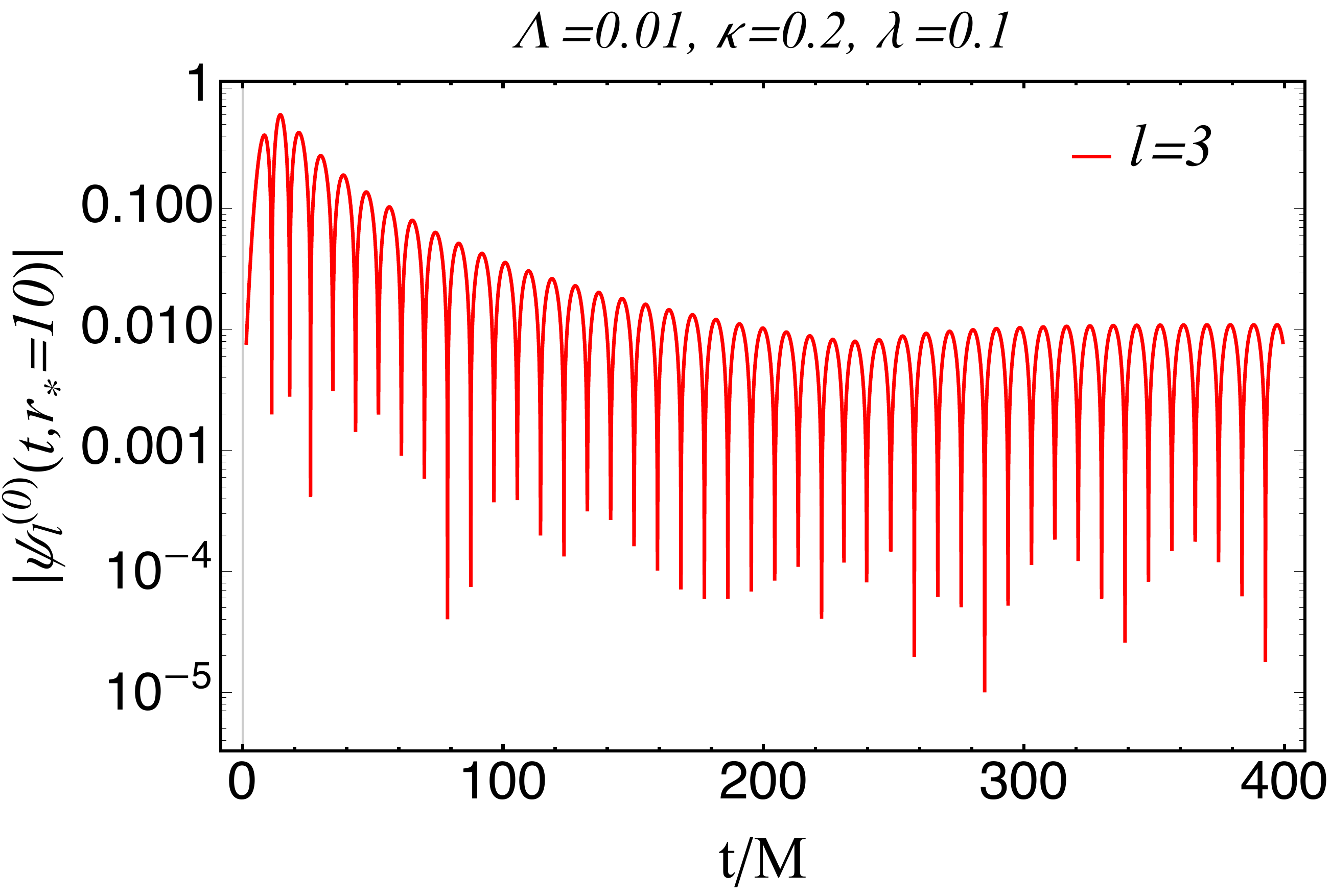}
	\endminipage
 \hfill
	%%%%%%%%%%%%%%%%%%%%%%%%
	\minipage{0.33\textwidth}
	\includegraphics[width=\linewidth]{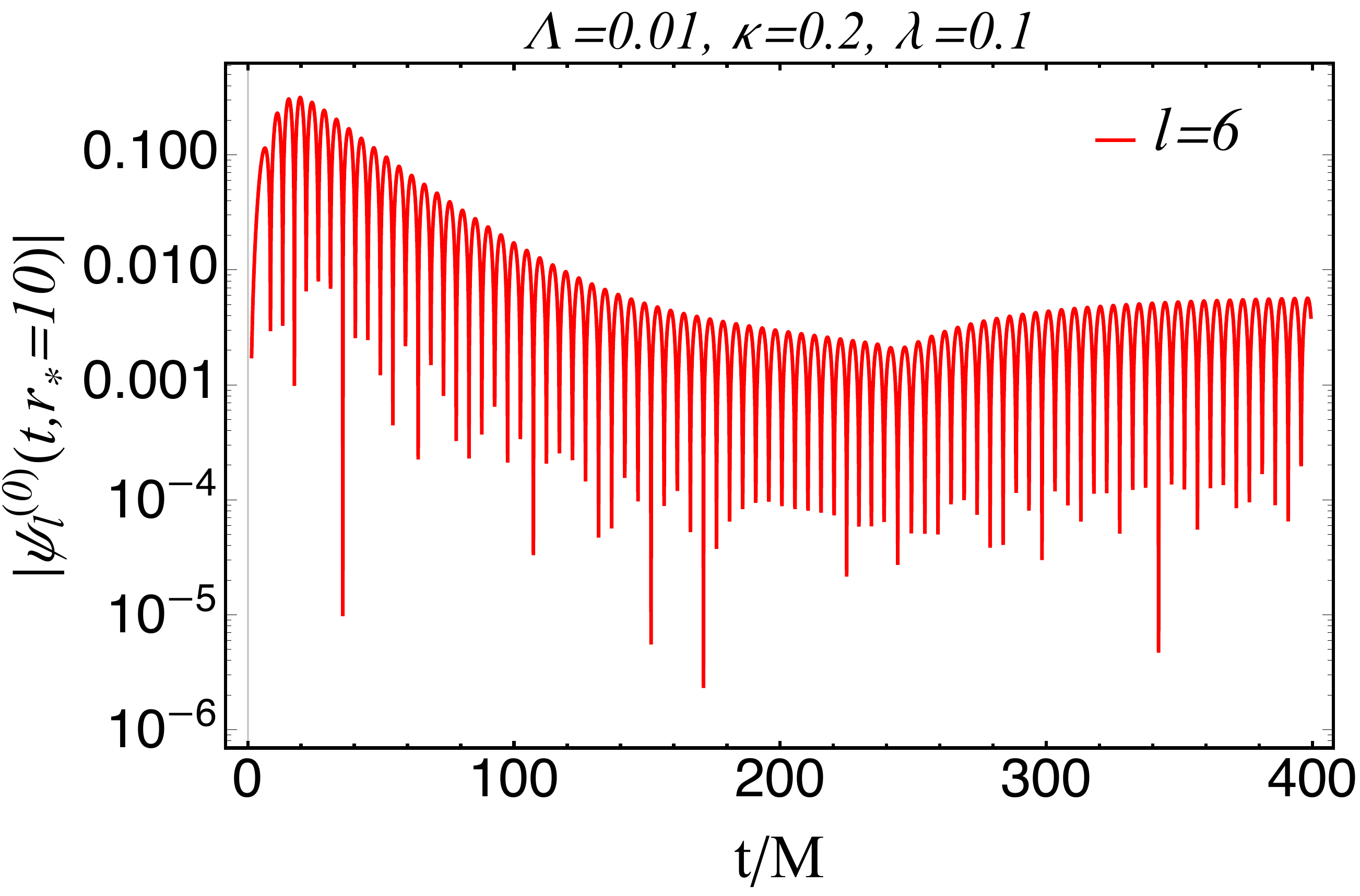}
	\endminipage\hfill
 \minipage{0.33\textwidth}
	\includegraphics[width=\linewidth]{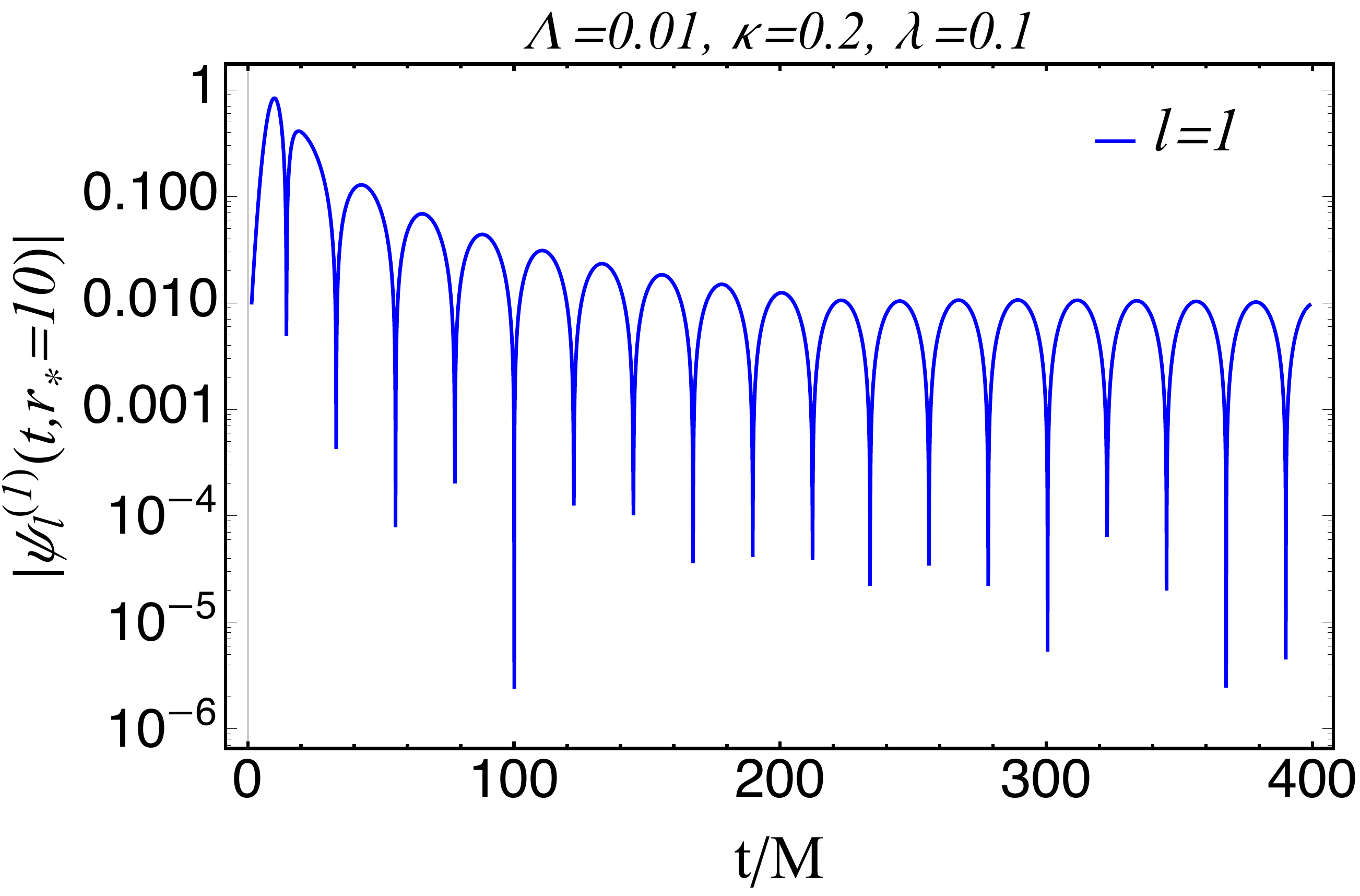}
	\endminipage\hfill
	%%%%%%%%%%%%%%%%%%%%%%%%
	\minipage{0.33\textwidth}
	\includegraphics[width=\linewidth]{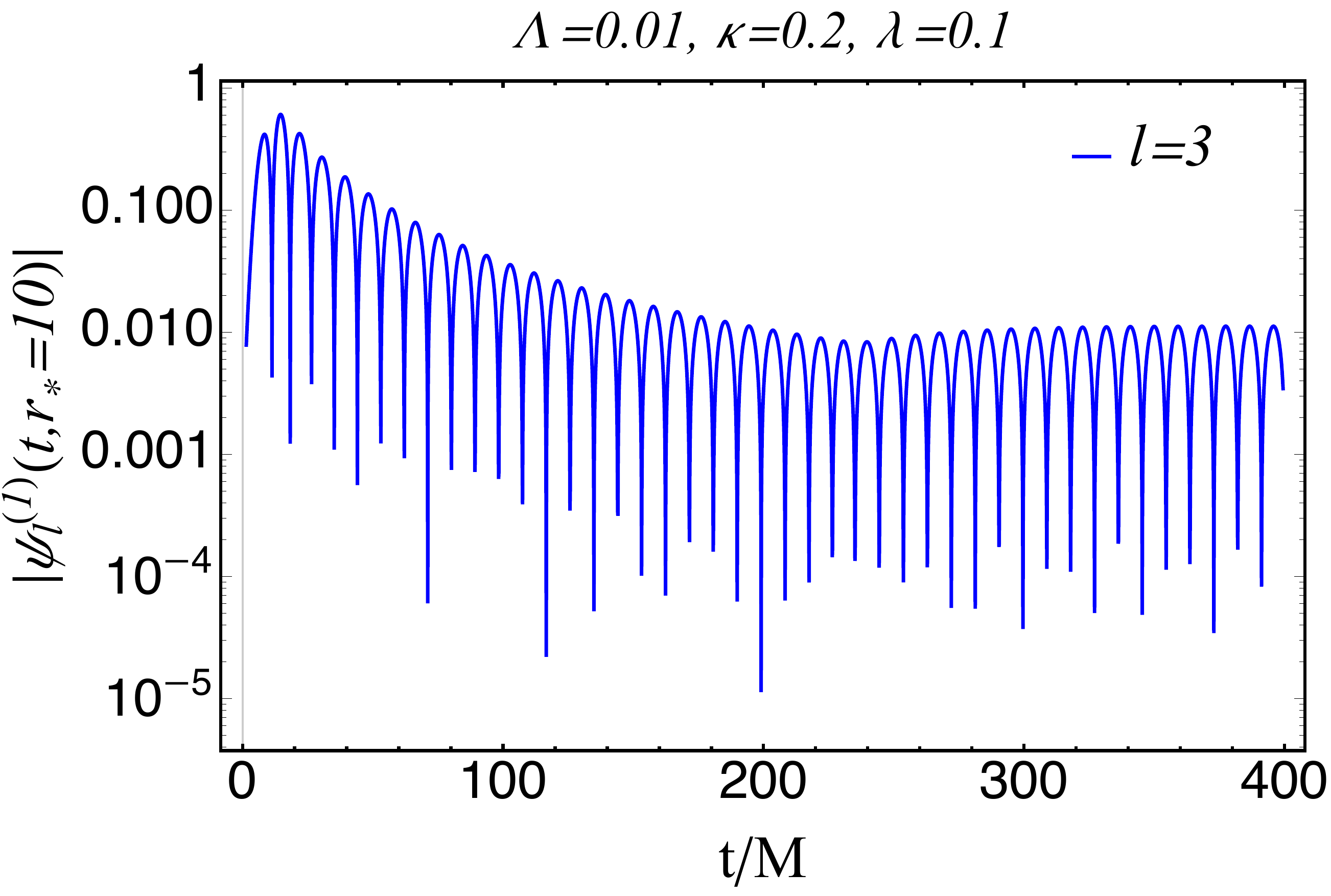}
	\endminipage
 \hfill
	%%%%%%%%%%%%%%%%%%%%%%%%
	\minipage{0.33\textwidth}
	\includegraphics[width=\linewidth]{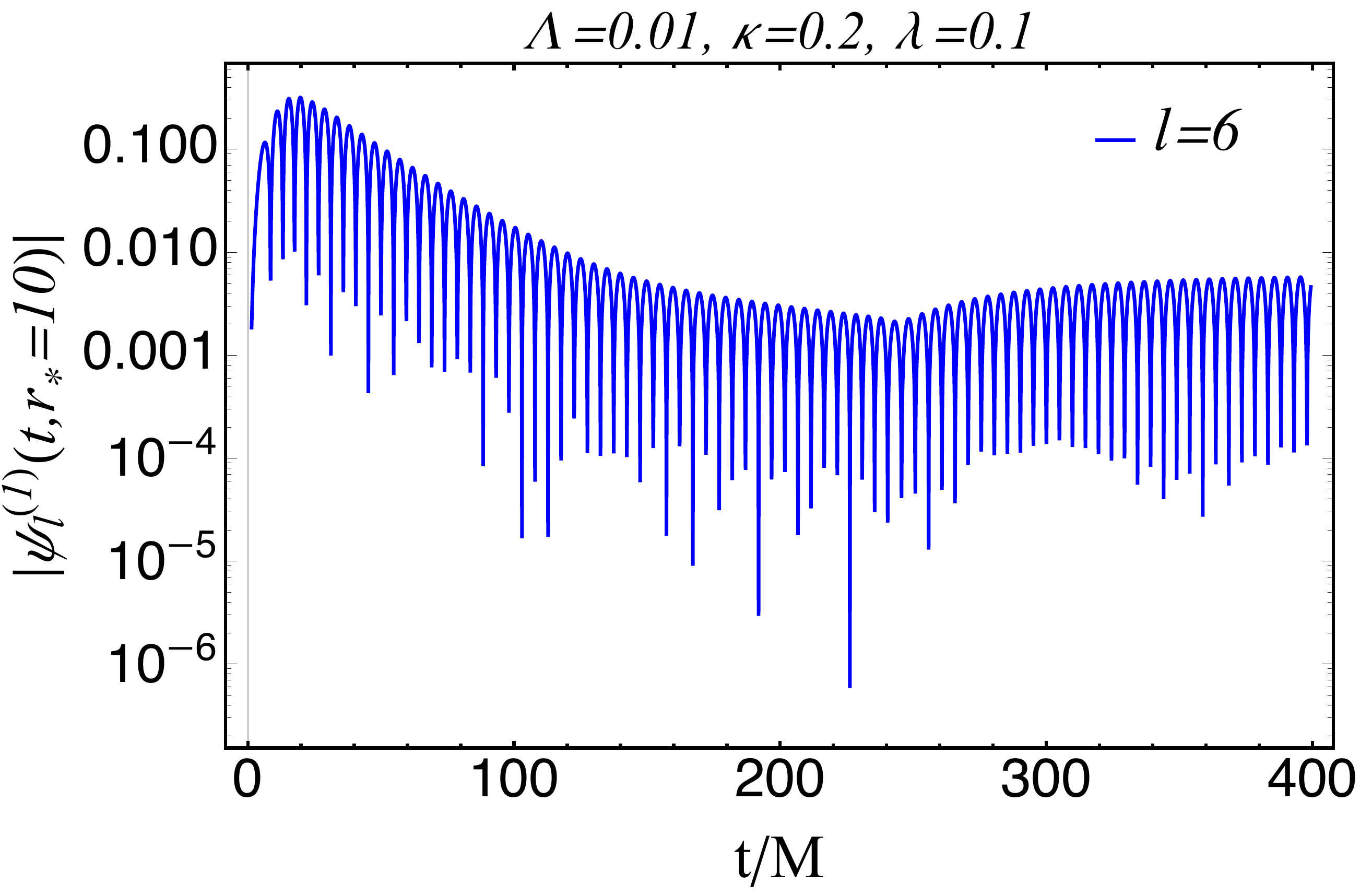}
	\endminipage
	\caption{Upper panel: Time evolution of scalar perturbation in the wormhole background for different values of $l$. Lower panel: The same for electromagnetic perturbation. In both of these cases, we consider $M=1$, $\Lambda=0.01$, $\kappa=0.2$, and $\lambda=0.1$.}\label{fig_latetime_different_lvalue}
\end{figure*}	
%%%%%%%%%%%%%%%%%%%%%%%%%%%%%%%%%%%%%%%%%%%%%%%%%%%%%%%%%%%%%%%%%
%%%%%%%%%%%%%%%%%%%%%%%%%%%%%%%%%%%%%%%%%%%%%%%%%%%%%%%%%%%%%%%%%
\begin{figure*}[htb!]
	%%%%%%%%%%%%%%%%%%%%%%%%
	\centering
	\minipage{0.33\textwidth}
	\includegraphics[width=\linewidth]{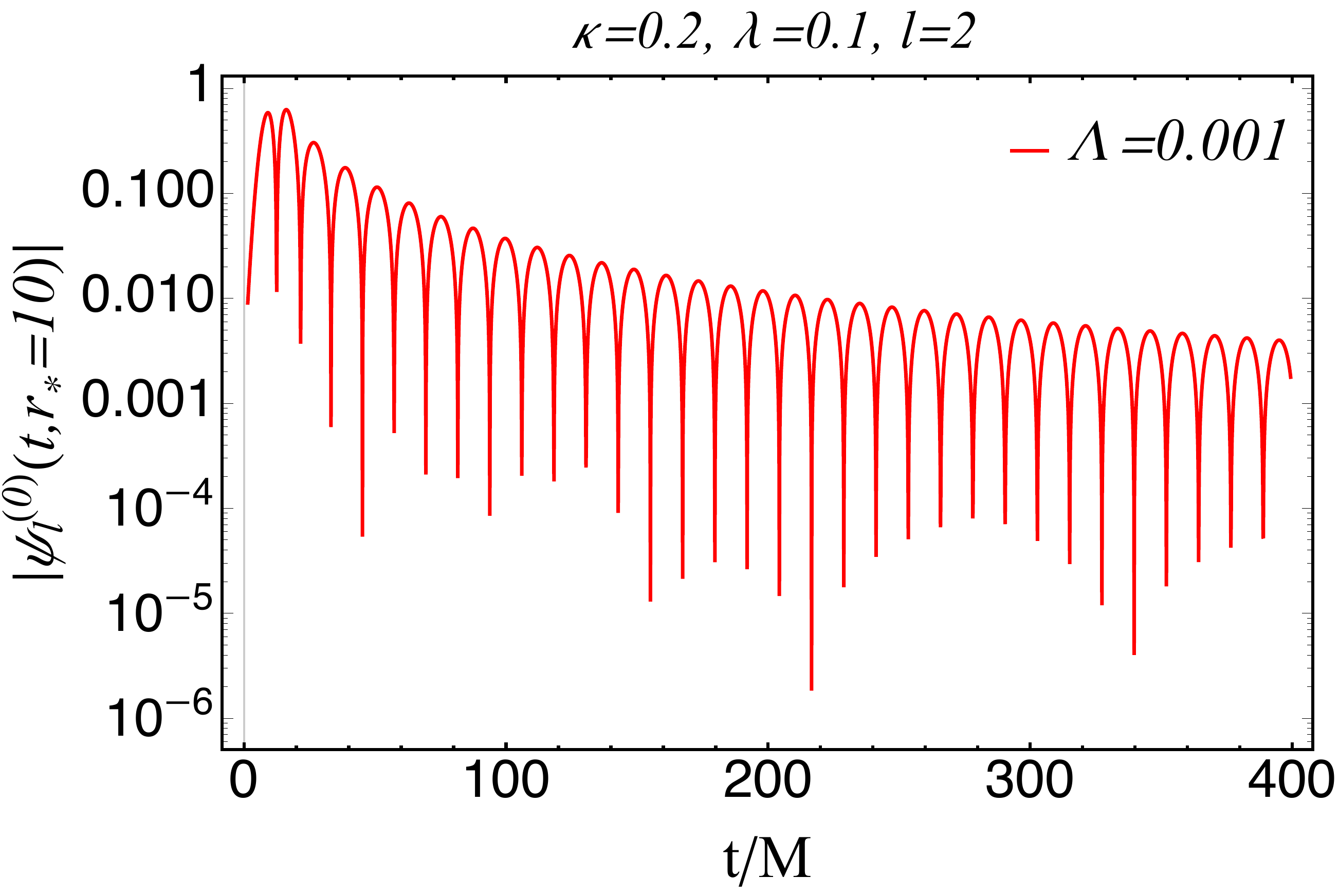}
	\endminipage\hfill
	%%%%%%%%%%%%%%%%%%%%%%%%
	\minipage{0.33\textwidth}
	\includegraphics[width=\linewidth]{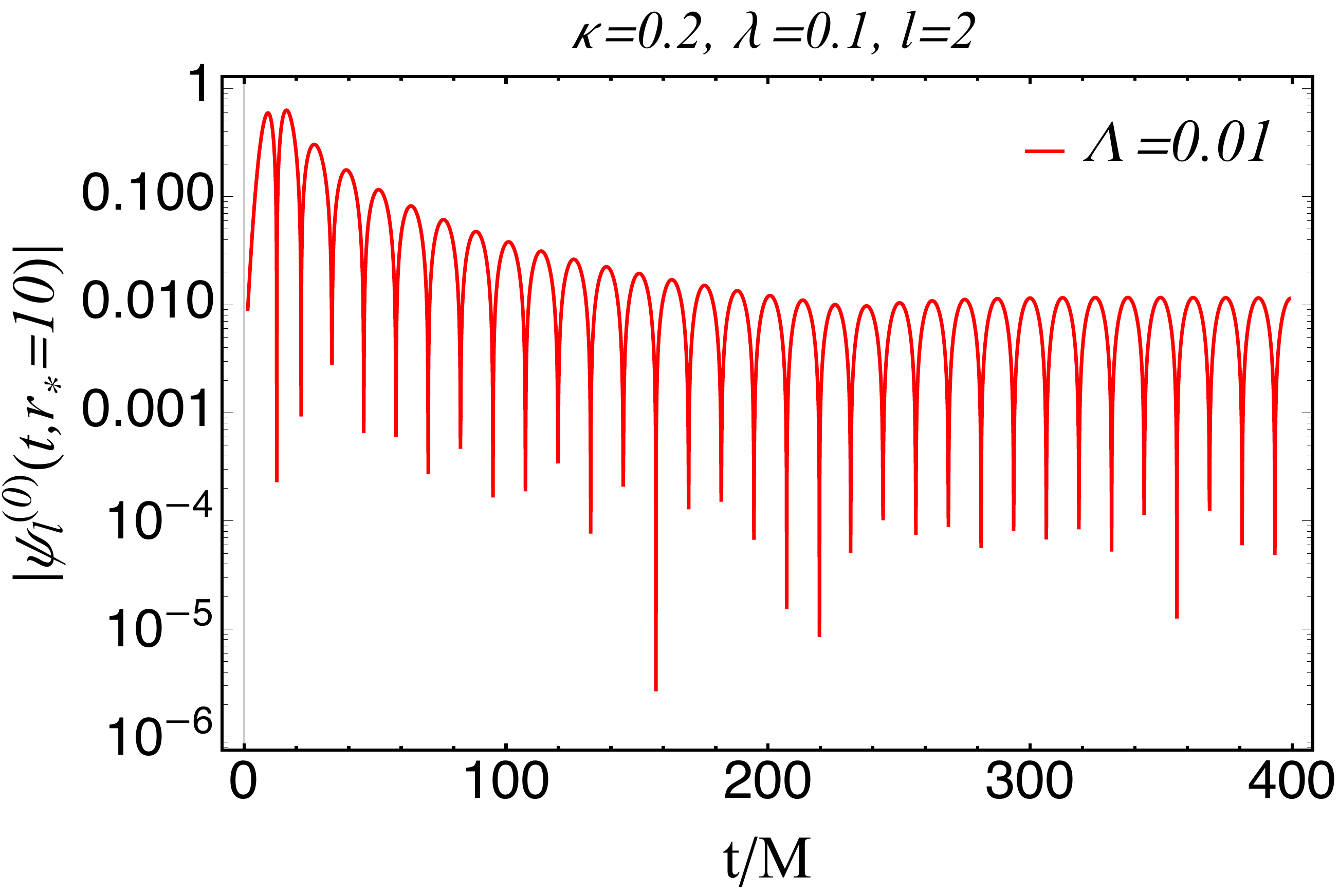}
	\endminipage
 \hfill
	%%%%%%%%%%%%%%%%%%%%%%%%
	\minipage{0.33\textwidth}
	\includegraphics[width=\linewidth]{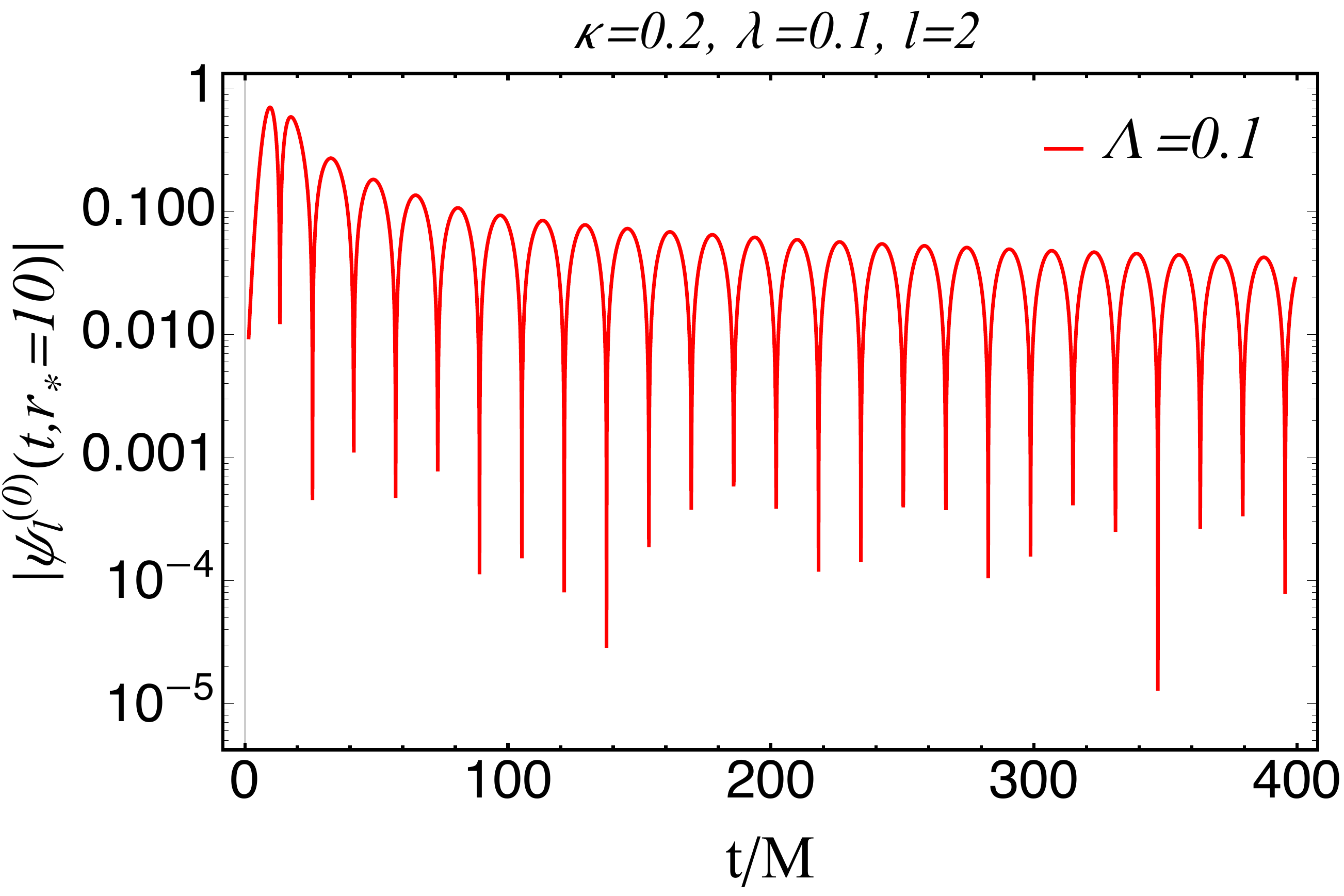}
	\endminipage\hfill
 \minipage{0.33\textwidth}
	\includegraphics[width=\linewidth]{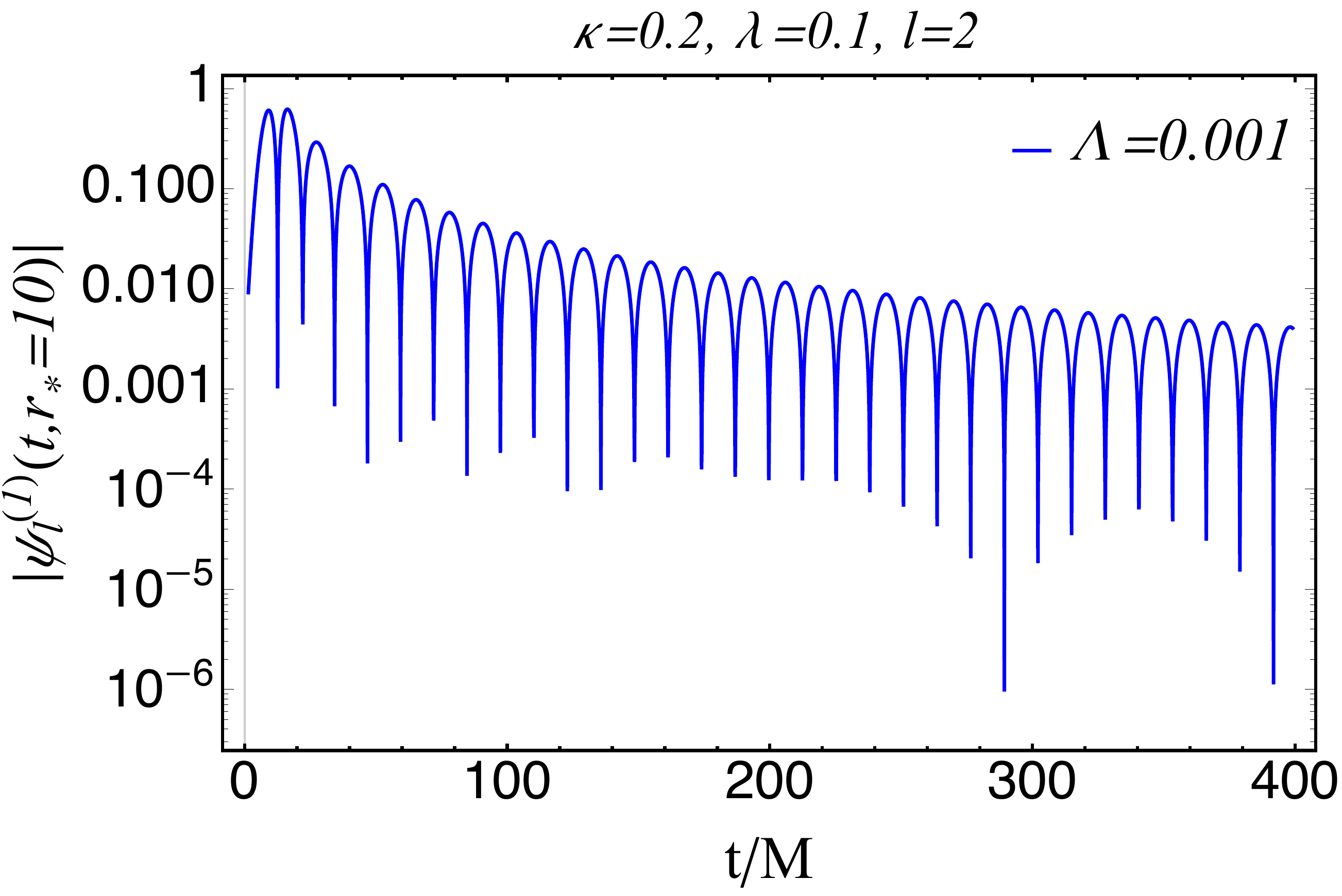}
	\endminipage\hfill
	%%%%%%%%%%%%%%%%%%%%%%%%
	\minipage{0.33\textwidth}
	\includegraphics[width=\linewidth]{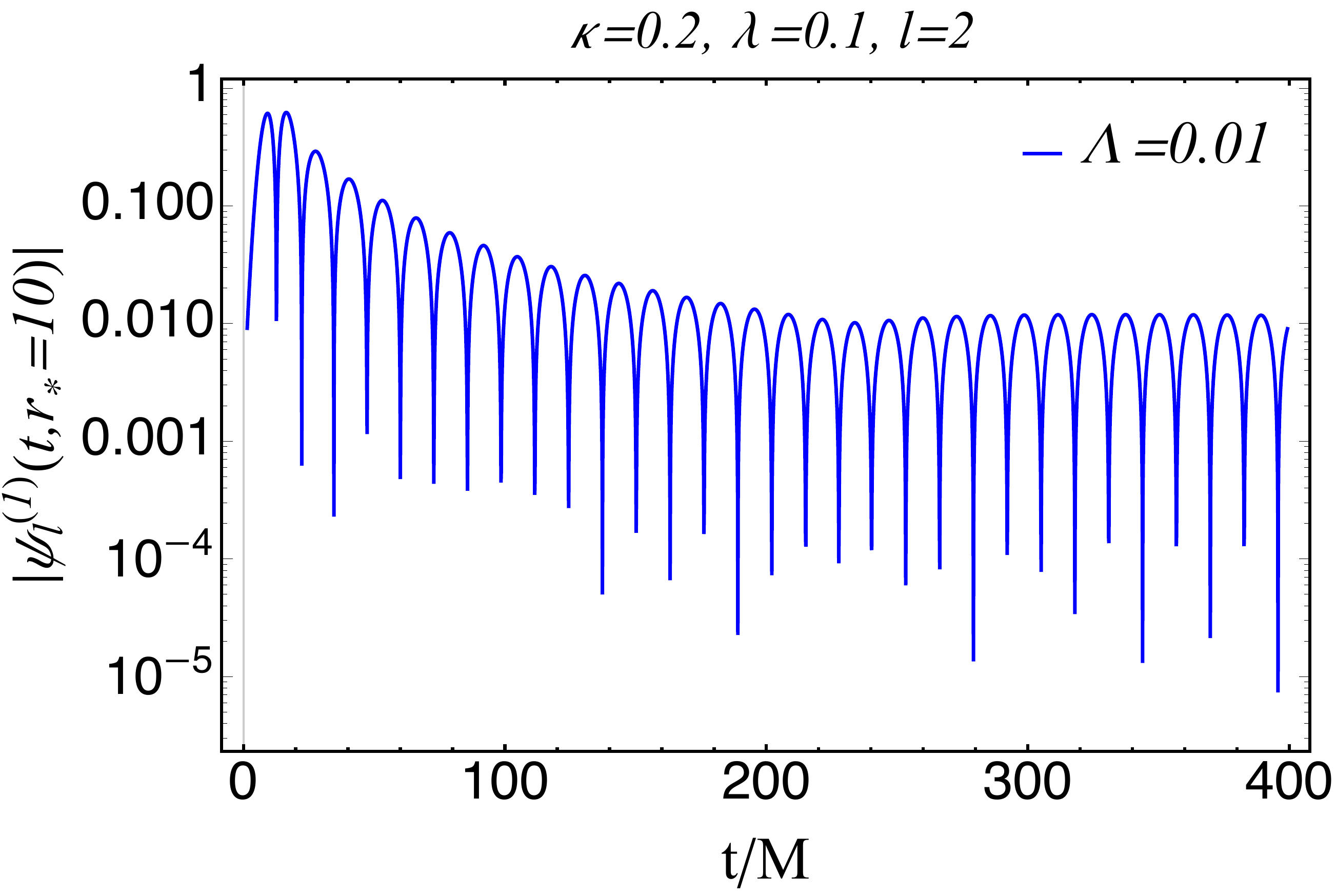}
	\endminipage
 \hfill
	%%%%%%%%%%%%%%%%%%%%%%%%
	\minipage{0.33\textwidth}
	\includegraphics[width=\linewidth]{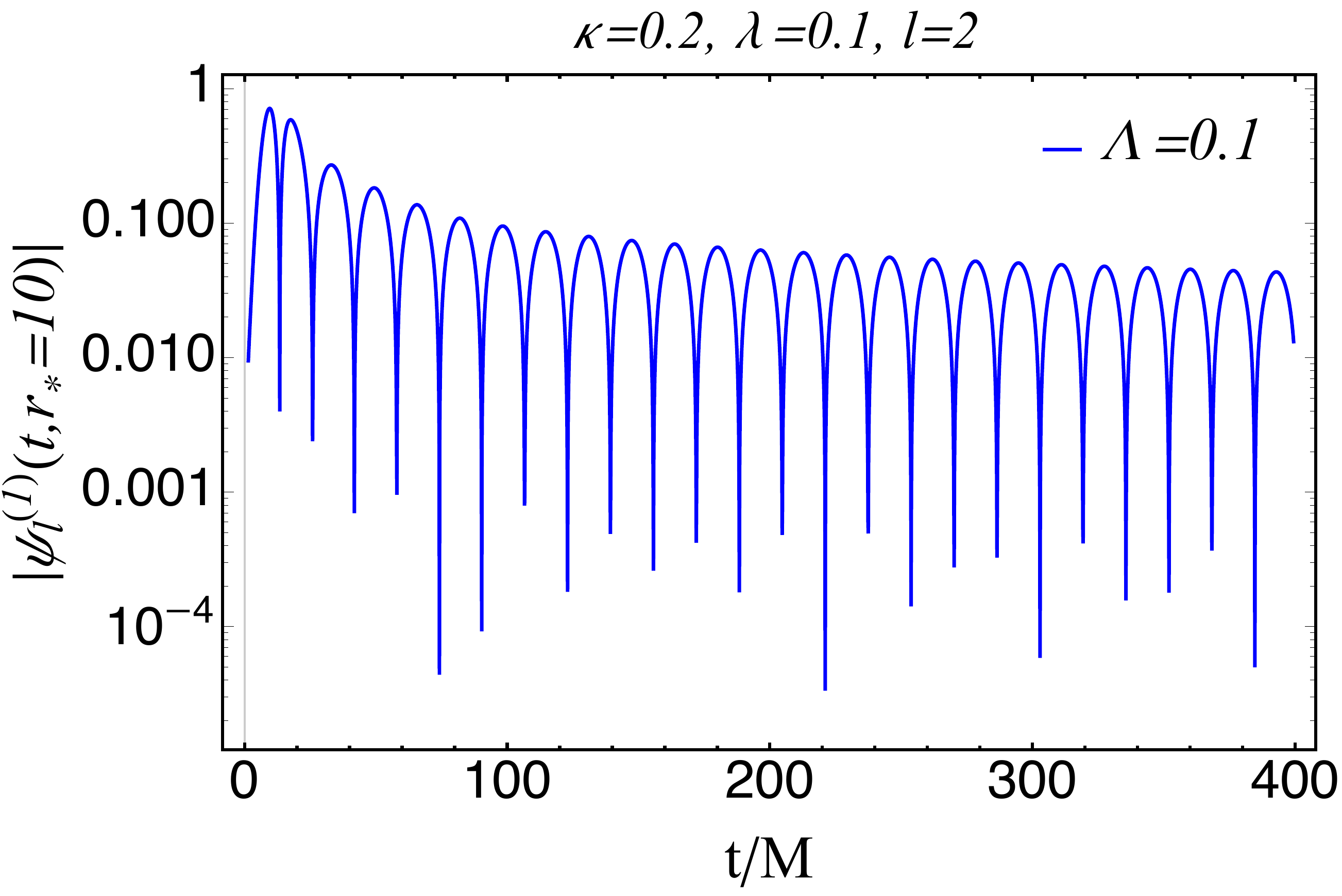}
	\endminipage
	\caption{Upper panel: Time evolution of scalar perturbation in the wormhole background for different values of $\Lambda$. Lower panel: The same for electromagnetic perturbation. In both of these cases, we consider $M=1$, $\kappa=0.2$, $\lambda=0.1$ and $l=2$.}\label{fig_latetime_different_lambvalue}
\end{figure*}	
%%%%%%%%%%%%%%%%%%%%%%%%%%%%%%%%%%%%%%%%%%%%%%%%%%%%%%%%%%%%%%%%%
%%%%%%%%%%%%%%%%%%%%%%%%%%%%%%%%%%%%%%%%%%%%%%%%%%%%%%%%%%%%%%%%%%%%%%%%%%%%%%%%%%%%%%%%%%%%%%%%%%%
%%%%%%%%%%%%%%%%%%%%%%%%%%%%%%%%%%%%%%%%%%%%%%%%%%%%%%%%%%%%%%%%%	
%%%%%%%%%%%%%%%%%%%%%%%%%%%%%%%%%%%%%%%%%%%%%%%%%%%%%%%%%%%%%%%%%
%%%%%%%%%%%%%%%%%%%%%%%%%%%%%%%%%%%%%%%%%%%%%%%%%%%%%%%%%%%%%%%%%%%%%%%%%%%%%%%%%%%%%%%%%%%%%%%%%%%
\section{Quasi-resonance and time-evolution of the fields}\label{Sec:QRM}
%%%%%%%%%%%%%%%%%%%%%%%%%%%%%%%%%%%%%%%%%%%%%%%%%%%%%%%%%%%%%%%%%
%%%%%%%%%%%%%%%%%%%%%%%%%%%%%%%%%%%%%%%%%%%%%%%%%%%%%%%%%%%%%%%%%
%Quasinormal modes represent characteristic oscillations in dissipative systems like black holes and wormholes. By introducing perturbations to their spacetime, either through additional fields or metric perturbations, these systems emit exponentially decaying sinusoids until reaching a stationary state. Intriguingly, in asymptotically flat spacetimes, the quasinormal modes associated with massive fields decay at a slower rate compared to their massless counterparts. The decay rate decreases as the mass term increases. This phenomenon, initially observed in the Reissner-Nordström black hole background for a massive scalar field, is referred to as quasi-resonances. It has been demonstrated by Konoplya and Zhidenko that such quasi-resonances are characteristic of perturbed spacetimes where the perturbing potential does not vanish at least at one of the boundaries.
In this section, we discuss the evolution of the scalar and electromagnetic field in the wormhole background. Note that the scalar and electromagnetic perturbation equations can be recasted in the following manner
%%%%%%%%%%%%%%%%%%%%%%%%%%%%%%%%%%
\begin{align}\label{Radial-black}
\frac{d^{2}\psi_{lm}^{(s)}}{dr_{*}^{2}}+[\omega^{2}-V_{l}^{(s)}(r)]\psi_{lm}^{(s)}=0~,
\end{align}  
%%%%%%%%%%%%%%%%%%%%%%%%%%%%%%%%%%
where, the perturbation potential can be written as 
%%%%%%%%%%%%%%%%%%%%%%%%%%%%%%%%%%
\begin{equation}
   V_{l}^{(s)}(r)=   e^{2a} \frac{l(l+1) }{r^2}+(1-s^2)\frac{ e^{(a-b)}}{r}\frac{d \left(e^{(a-b)}\right)}{dr}~.
\end{equation}
 Here, $s=0$ and $s=1$ correspond to scalar and electromagnetic perturbation, respectively. As discussed above, we obtain the tortoise coordinate $r_*$ by solving the differential equation \autoref{tortoise} numerically with the condition that $r_*=0$ at the throat $r=\rth$. Note that both the scalar and electromagnetic perturbation potential is symmetric about the wormhole throat $r_*=0$. Thus, the $\psi_{lm}^{(s)}$ should be either an even (symmetric) or an odd (antisymmetric) function of $r_*$. In this study, we have chosen $\psi_{lm}^{(s)}(t,r_*)$ to be an odd function. Interestingly, the perturbation potential $V_{l}^{(s)}(r)$ is non-vanishing both at the throat $\rth$ and the cosmological horizon $r_c$. As a consequence, arbitrarily long-long lived modes (modes with purely real frequency), so-called the \textit{quasi-resonance} modes, can exist in this spacetime \cite{Konoplya:2004wg}.  The occurrence of these quasi-resonances was initially observed within the context of the Reissner-Nordström black hole background, for a massive scalar field perturbation \cite{Ohashi:2004wr}. Such phenomena, have been subsequently demonstrated by Konoplya and Zhidenko to be a characteristic of perturbed spacetimes where the perturbing potential does not vanish at least at one of the boundaries \cite{Konoplya:2004wg}. To understand this, we introduce two parameters
 %%%%%%%%%%%%%%%%%%%%%%%%%%%%%%%%%%
\begin{equation}
\begin{aligned}
    \omega_{t}^{(s)}=\sqrt{\omega^2- V_{l}^{(s)}(\rth)} 
\end{aligned}
\end{equation}
and 
\begin{equation}
\begin{aligned}
    \omega_{c}^{(s)}=\sqrt{\omega^2- V_{l}^{(s)}(r_c)} 
\end{aligned}
\end{equation}
where 
\begin{equation}
\begin{aligned}
   V_{l}^{(s)}(\rth)&= \kappa^2 \frac{l(l+1) }{\rth^2}+(1-s^2)\left[\frac{ e^{(a-b)}}{r}\frac{d \left(e^{(a-b)}\right)}{dr}\right]_{\rth}~,\\
   V_{l}^{(s)}(r_c)&= \kappa^2 \frac{l(l+1) }{r_c^2}+(1-s^2)\left[\frac{ e^{(a-b)}}{r}\frac{d \left(e^{(a-b)}\right)}{dr}\right]_{r_c}~
\end{aligned}
\end{equation}
 %, i.e., both $\left(\ps modesi_{lm}^{(s)}(r_*)+\psi_{lm}^{(s)}(-r_*)\right)$ and $\left(\psi_{lm}^{(s)}(r_*)-\psi_{lm}^{(s)}(-r_*)\right)$ is a solution of \autoref{Radial-black}.
For improved readability, we will omit the superscript $(s)$ in $\psi_{lm}^{(s)}$, $V_{l}^{(s)}$, $\omega_{t}^{(s)}$, and $\omega_{c}^{(s)}$ for analysis below, as it is applicable for both the scalar and electromagnetic fields. The solution of the perturbation equation \autoref{Radial-black} near the wormhole throat ($r\to\rth$) can be written as
\begin{equation}\label{BC1}
\begin{aligned}
   \psi_{lm}&=A_t e^{i\omega_t r_*}+B_t e^{-i\omega_t r_*}~\\
   &=C_t \sin\left(\omega_t r_*\right)
\end{aligned}
\end{equation}
where, $A_t$ and $B_t$ are two arbitrary constants and $C_t=2i A_t=-2i B_t$. Here, we use the fact that the wave function is a odd function. Similarly, near the cosmological horizon ($r\to r_c$), the the solution of the perturbation equation \autoref{Radial_em} can be written as 
\begin{equation}\label{BC2}
\begin{aligned}
   \psi_{lm}&=A_c e^{i\omega_c r_*}+B_c e^{-i\omega_c r_*}~
\end{aligned}
\end{equation}
Assuming no ingoing wave from the cosmological horizon, we set $B_c=0$. Given the boundary conditions \autoref{BC1} and \autoref{BC2}, it is straightforward to demonstrate the existence of modes with purely real frequencies. We start by multiplying \autoref{Radial-black} by $\overline\psi_{lm}$ and integrating in the range $r_*\in \left(x_t,x_c\right)$, which yields
\begin{equation}
    \begin{aligned}
        \int_{x_t}^{x_c}dr_* \overline\psi_{lm}\left[\frac{d^{2}\psi_{lm}}{dr_{*}^{2}}+[\omega^{2}-V_{l}^{(s)}(r)]\psi_{lm}\right]=0~,
    \end{aligned}
\end{equation}
where, $x_c=r_*(r_c)$ and $x_t=r_*(\rth)$. After performing integration by parts, we arrive at
\begin{equation}
     \begin{aligned}\label{condi1}
         \int_{x_t}^{x_c}dr_* \bigg[\left(\omega^{2}-V_{l}^{(s)}(r)\right)|\psi_{lm}|^2 &-\left|\partial_{r_*}\psi_{lm}\right|^2\bigg]\\&+\overline\psi_{lm} \partial_{r_*}\psi_{lm}\bigg|_{x_t}^{x_c}=0~,
    \end{aligned}
\end{equation}
Using \autoref{BC1} and \autoref{BC2}, we obtain 
\begin{equation}\label{condi2}
    \begin{aligned}
\overline\psi_{lm} \partial_{r_*}\psi_{lm}\bigg|_{x_t}^{x_c}=i \omega_c |A_t|^2~e^{-2\textrm{Im}(\omega_c)x_c}~,
    \end{aligned}
\end{equation}
Substituting \autoref{condi2} into \autoref{condi1} and separating real and imaginary parts, we obtain
\begin{equation}\label{condi3}
    \begin{aligned}
 \textrm{Re}(\omega_c)|A_t|^2=0~,
    \end{aligned}
\end{equation}
which implies that there is no dissipation of energy through the cosmological horizon as $\textrm{Re}(\omega_c)=0$ \cite{Konoplya:2004wg}. As a result, the oscillating modes are not decaying, a phenomena analogous to standing waves on a fixed string. Since $\omega_c$ is imaginary; this also bounds the quasi-resonant frequencies through the following relation
\begin{equation}\label{quasi_resonant_condition}
    \begin{aligned}
 \left|\omega_{\textrm{QRM}}\right|&<\sqrt{V_{l}(r_c)}\\&=\sqrt{\kappa^2 \frac{l(l+1) }{r_c^2}+(1-s^2)\left[\frac{ e^{(a-b)}}{r}\frac{d \left(e^{(a-b)}\right)}{dr}\right]_{r_c}}
    \end{aligned}
\end{equation}
As mentioned previously, a similar phenomenon occurs in asymptotically flat black holes in the presence of a massive field \cite{Ohashi:2004wr, Konoplya:2004wg, Konoplya:2005hr, Konoplya:2006br, Zhidenko:2006rs, Churilova:2019qph, Hod:2015goa, Hod:2016jqt}. In this scenario, the perturbation potential does not vanish at asymptotic infinity but approaches a finite value $\mu^2$, where $\mu$ corresponds to the mass of the field. Consequently, the frequencies of quasi-resonant modes are bounded by the mass of the field. Furthermore, the decay rate of these quasi-resonant modes decreases as the mass term increases. In fact, for a certain value of $\mu$, the decay rate of the fundamental modes reaches zero. If the mass of the field is further increased, even the higher overtones of the quasi-resonant modes cease to decay.\par 
Interestingly, when considering massive scalar fields in asymptotically de Sitter black holes, no such phenomena can be observed. This is due to the fact that the perturbation potential becomes zero at the cosmological horizon (see \autoref{quasi_resonant_condition}) \cite{Konoplya:2004wg, Chang:2007zzv}. However, for the wormhole under consideration, these quasi-resonant modes exist even when the perturbing field is massless. In this case, $\sqrt{V_{l}(r_c)}$ serves as an effective mass term. Notably, the behavior of the term primarily depends on three parameters: $\kappa$, $l$, and $\Lambda$ (through $r_c$), with $\sqrt{V_{l}(r_c)}$ increasing as these parameters increase. To investigate whether the decay rate of quasinormal modes decreases with the increase of these parameters, we examine the time-evolution of the fields.\par  
 In order to obtain the time-domain behavior of the fields, we use inverse Fourier transformation $-i\omega \psi_{lm}^{(s)}=\partial_t \psi_{lm}^{(s)}$ in \autoref{Radial-black} and then rewriting the equation in terms of advanced ($v=t+r_{*}$) and retarded time ($u=t-r_{*}$). The resulting equation $4\partial_u\partial_v \psi_{lm}^{(s)}=-V_{l}^{(s)}\psi_{lm}^{(s)}$ can be discretized in the following manner 
 %%%%%%%%%%%%%%%%%%%%%%%%%%%%%%%%%%
\begin{align}\label{discretization}
\psi(N)=\psi(W)+\psi(E)-\psi(S)-\frac{h^2}{8}V(s)\left(\psi(W)+\psi(E)\right)
\end{align}  
%%%%%%%%%%%%%%%%%%%%%%%%%%%%%%%%%%
where, $h$ is the grid scale factor and $N=(u+h,v+h)$, $W=(u+h,v)$, $E=(u,v+h)$ and $S=(u,v)$. Here, we have omitted the subscript and the superscript of the master function and the perturbation potential for the sake of brevity. Furthermore, we impose the following initial condition
%%%%%%%%%%%%%%%%%%%%%%%%%%%%%%%%%%
\begin{equation}\label{initial_conditions}
  \begin{aligned}
\psi(u,0)=\exp\left[\frac{-{(u-10)^2}}{18}\right]\,,\quad \psi(0,v)=1~.
\end{aligned}    
\end{equation}
%%%%%%%%%%%%%%%%%%%%%%%%%%%%%%%%%%
The figures \autoref{fig_latetime_different_kappa},  \autoref{fig_latetime_different_lvalue} and \autoref{fig_latetime_different_lambvalue} illustrate the time-domain signal of the scalar and electromagnetic perturbation. Several notable features can be observed from these signals:
\begin{itemize}
    \item There is initial decay period, after which the fields ceased  to decay.
    \item For smaller values of $\kappa$, $l$ and $\Lambda$, the initial decay period is significantly longer.
    \item By increasing the values of the parameters $\kappa$, $l$ and $\Lambda$, the duration of the decay period becomes shorter. At sufficiently late times, we observe the emergence of extremely long-lived modes. 
\end{itemize}
%%%%%%%%%%%%%%%%%%%%%%%%%%%%%%%%%%%%%%%%%%%%%%%%%%%%%%%%%%%%
These observations are directly connected to the phenomenon of quasi-resonance described earlier. The effective mass term, represented by $\sqrt{V_{l}(r_c)}$, depends on the parameters $\kappa$, $l$ and $\Lambda$. As these parameters increase, the effective mass term also increases. Note that, at sufficiently late times, the signal is dominated by the longest-lived mode, which corresponds to the fundamental mode. The decay rate of this fundamental mode decreases as the effective mass term increases. Consequently, in the time-domain signal, we observe an almost non-decaying mode at later times. Additionally, even the decay rate of the higher overtones decreases with an increase in the effective mass term. This accounts for the shorter duration of the initial decay period as the parameters $\kappa$, $l$ and $\Lambda$ increase.
%This observation is intrinsically linked to the phenomena of quasi-resonance described above. The effective mass term $\sqrt{V_{l}(r_c)}$ depends on the parameters $\kappa$, and $l$, with the value of the effective mass increases with the increase of these parameters. Note that at sufficiently late the the signal is dominated by the longest-lived mode, i.e., the fundamental mode, the decay rate of which decreases with increase of the effective mass term. This phenomena is observed in the time-domain signal, where we find almost non-decaying mode at sufficiently later times. Furthermore, even the decay rate of the overtones decreases with the increase of the effective mass term. That is why we see the duration of decay period decreases with the increase of the paramater $\kappa$ and $l$. 
%In particular, \autoref{fig_latetime_different_kappa} displays the time evolution of the scalar and electromagnetic field for $M=1$, $\Lambda=0.01$, $\lambda=0.1$, $l=2$ and different values of $\kappa$.
%As can be seen from the plots, after a initial decay period the fields cease to decay. Furthermore, we increase the value of the parameter $\kappa$, the duration of the decay period get shortened. The same conclusion can be drawn if we increase the value of $l$ (see \autoref{fig_latetime_different_lvalue}). This is intrinsically linked with phenoemna of quasi-resonance. The effect mass term  $\sqrt{V_{l}(r_c)}$ increases with the increase of the parameter $\kapp$ and $l$.  This is in accordance to the phenomena described above. \par
 \section{Conclusion}\label{Sec:Conclusion}
%%%%%%%%%%%%%%%%%%%%%%%%%%%%%%%%%%%%%%%%%%%%%%%%%%%%%%%%%%%%%%%%%%%%%%%%%%%%%%%%%%%%%%%%%%%%%%%%%%%
The bi-gravity theory involves two interacting gravitons, one massive and the other massless, and is described by two dynamical metrics $g_{\mu\nu}$ and $f_{\mu\nu}$. This ghost-free theory offers an alternative to Einstein's general relativity and has been the subject of recent discussions in both cosmological and astrophysical contexts. The known solutions in this theory can be classified into three types \cite{Volkov:2013roa}. The first type comprises solutions where the two metrics are proportional to each other within the same frame. The second type involves non-diagonal metrics in a spherically symmetric setup. The third type encompasses bi-diagonal metrics in the same coordinate system that are not proportional to each other. In this paper, we report the finding of a new exact vacuum solution, which belongs to the class of bi-gravity theories with $\beta_{2}=\beta_{3} =0$ and falls into the third category mentioned earlier. Interestingly, we find that the Killing horizon of the metrics $g_{\mu\nu}$ and $f_{\mu\nu}$ (if they exist!) coincides with each other and, thus, consistent with Deffayet and Jacobson's theorem \cite{Deffayet:2011rh}. The solution exhibits spatial Schwarzschild-de Sitter geometry, with the parameter $\beta_1 m^2$ serving as a cosmological constant. The metrics also depend on two additional parameters, $\kappa$ and $\lambda$. Notably, for the choices $\kappa=0$ and $\lambda=1$, the solution resembles a Schwarzschild-de Sitter black hole. However, for the choices $\kappa^2\neq 0$ and  $\lambda=0$ or $\kappa^2>\lambda^2$, the solution describes a wormhole with the position of the wormhole throat corresponds to the smallest positive solution of the equation $1-2M/r-\Lambda r^2/3=0$.  We have verified that the weak energy condition is violated near the wormhole throat and thus, the flare-out condition is satisfied.\par
A similar type of wormhole solution (with $\Lambda=0$) was previously reported in \cite{Dadhich:2001fu} within the framework of general relativity. In that context, the solution was obtained by imposing the restriction $\rho=\rho_t=0$ on the energy-momentum tensor, where $\rho$ represents the density measured by a static observer and $\rho_t$ represents the convergence density experienced by a timelike congruence. Subsequently, it was discovered that such wormholes can also exist in the context of brane-world scenarios. Interestingly such solution can be obtained in the bi-gravity theory for the choice of parameters $\beta_{1} = \beta_{4}$. 
\par
Furthermore, we investigate the behavior of scalar and electromagnetic fields in the wormhole spacetime. Remarkably, we observe long-lived modes known as quasi-resonances in the quasinormal spectrum. This phenomenon is particularly prominent in the quasinormal spectrum of massive fields \cite{Ohashi:2004wr, Konoplya:2004wg, Konoplya:2005hr, Konoplya:2006br, Zhidenko:2006rs, Churilova:2019qph, Hod:2015goa, Hod:2016jqt}. It has been recognized as a general feature when the perturbation potential of the associated master equation of the field does not vanish at one of the boundaries. In this study, we analytically and numerically demonstrate the presence of such quasi-resonant modes in the wormhole spacetime considered here, even for massless fields. Additionally, we find that the decay rate of these quasi-resonant modes decreases as the parameters $\kappa$, $l$ and $\Lambda$ increase.\par
An interesting extension of our study would involve analyzing the characteristics of null and time-like geodesics in the wormhole spacetime. Exploring the presence of quasi-resonant modes induced by gravitational perturbations would also be of great interest. Additionally, it would be interesting to explore the existence of rotating wormhole solutions within the framework of bi-gravity theories. However, these topics are beyond the scope of this paper and will be addressed in our future work.
\vspace{5mm}
	%%%%%%%%%%%%%%%%%%%%%%%%%%%%%%%%%%%%%%%%%%%%%%%%%%%%%%%%%%%%%%%%%%%%%%%%%%%%%%%%%%%%%%%%%%%%%%%%%%%
	%%%%%%%%%%%%%%%%%%%%%%%%%%%%%%%%%%%%%%%%%%%%%%%%%%%%%%%%%%%%%%%%%%%%%%%%%%%%%%%%%%%%%%%%%%%%%%%%%%%
	%%%%%%%%%%%%%%%%%%%%%%%%%%%%%%%%%%%%%%%%%%%%%%%%%%%%%%%%%%%%%%%%%%%%%%%%%%%%%%%%%%%%%%%%%%%%%%%%%%%
	%%%%%%%%%%%%%%%%%%%%%%%%%%%%%%%%%%%%%%%%%%%%%%%%%%%%%%%%%%%%%%%%%%%%%%%%%%%%%%%%%%%%%%%%%%%%%%%%%%%
	\section*{Acknowledgements}
	   The authors like to thank Sayan Kar, Sumanta Chakraborty and Arpan Bhattacharyya for useful discussions. The research of M.R is supported by the National Post Doctoral Fellowship grant (Reg. No. PDF/2021/001234) by  SERB, Government of India. AAS acknowledges the funding from SERB, Govt of India under the research grant MTR/20l9/000599. SSB acknowledges the funding from the University Grants Commission, Govt of India under JRF scheme.
	%%%%%%%%%%%%%%%%%%%%%%%%%%%%%%%%%%%%%%%%%%%%%%%%%%%%%%%%%%%%%%%%%%%%%%%%%%%%%%%%%%%%%%%%%%%%%%%%%%%%%
	%%%%%%%%%%%%%%%%%%%%%%%%%%%%%%%%%%%%%%%%%%%%%%%%%%%%%%%%%%%%%%%%%%%%%%%%%%%%%%%%%%%%%%%%%%%%%%%%%%%%%
	%%%%%%%%%%%%%%%%%%%%%%%%%%%%%%%%%%%%%%%%%%%%%%%%%%%%%%%%%%%%%%%%%%%%%%%%%%%%%%%%%%%%%%%%%%%%%%%%%%%%%	%%%%%%%%%%%%%%%%%%%%%%%%%%%%%%%%%%%%%%%%%%%%%%%%%%%%%%%%%%%%%%%%%%%%%%%%%%%%%%%%%%%%%%%%%%%%%%%%%%%
	%%%%%%%%%%%%%%%%%%%%%%%%%%%%%%%%%%%%%%%%%%%%%%%%%%%%%%%%%%%%%%%%%%%%%%%%%%%%%%%%%%%%%%%%%%%%%%%%%%%
	%%%%%%%%%%%%%%%%%%%%%%%%%%%%%%%%%%%%%%%%%%%%%%%%%%%%%%%%%%%%%%%%%%%%%%%%%%%%%%%%%%%%%%%%%%%%%%%%%%%
	%%%%%%%%%%%%%%%%%%%%%%%%%%%%%%%%%%%%%%%%%%%%%%%%%%%%%%%%%%%%%%%%%%%%%%%%%%%%%%%%%%%%%%%%%%%%%%%%%%%
	%%%%%%%%%%%%%%%%%%%%%%%%%%%%%%%%%%%%%%%%%%%%%%%%%%%%%%%%%%%%%%%%%%%%%%%%%%%%%%%%%%%%%%%%%%%%%%%%%%%
	%%%%%%%%%%%%%%%%%%%%%%%%%%%%%%%%%%%%%%%%%%%%%%%%%%%%%%%%%%%%%%%%%%%%%%%%%%%%%%%%%%%%%%%%%%%%%%%%%%%
% \newpage
	\appendix
	
	%\labelformat{section}{Appendix #1}
	%	\labelformat{subsection}{Appendix #1}
	%	\labelformat{subsubsection}{Appendix #1}
	%%%%%%%%%%%%%%%%%%%%%%%%%%%%%%%%%%%%%%%%%%%%%%%%%%%%%%%%%%%%%%%%%%%%%%%%%%%%%%%%%%%%%%%%%%%%%%%%%%%
	%%%%%%%%%%%%%%%%%%%%%%%%%%%%%%%%%%%%%%%%%%%%%%%%%%%%%%%%%%%%%%%%%%%%%%%%%%%%%%%%%%%%%%%%%%%%%%%%%%%
	%%%%%%%%%%%%%%%%%%%%%%%%%%%%%%%%%%%%%%%%%%%%%%%%%%%%%%%%%%%%%%%%%%%%%%%%%%%%%%%%%%%%%%%%%%%%%%%%%%%
 	\section{Bi-diagonal solution in bi-metric theory}\label{App:Bi-diagonal}
  The spherically symmetric solutions of bi-metric gravity theory can categorized into two distinct groups, namely, the bi-diagonal solutions and the non-bi-diagonal solutions \cite{Volkov:2014ooa, Schmidt_May_2016}. As the nomenclature indicates, both the metrics $g_{\mu\nu}$ and $f_{\mu\nu}$ are diagonal for the solutions belonging to the former class, whereas, for the latter case, it is impossible to transform both metrics into a diagonal form simultaneously. 
  In this section, we investigate the properties of the static, spherically symmetric, vacuum solutions of \autoref{einstein_equation} under the assumption $\beta_2=\beta_3=0$ to check whether the solutions are bi-diagonalizable or not.  
%We are considering the spherical symmetric space-time in the Bi-metric gravity. there are two metrics $g_{\mu \nu}$ and $f_{\mu \nu}$. Both these metrics are defined by their equations of motion \autoref{field_equation} and \autoref{} respectively. 
We are studying the space-time manifold in a single local coordinate system. Note that, with the proper coordinate transformation, we can always make one of the metrics in the diagonal form. Thus, we start with the following ansatz for the metrics \cite{Volkov:2014ooa, Schmidt_May_2016}
%%%%%
\begin{equation}\label{general_spherical_metric}
    \begin{aligned} 
ds_{g}^{2} & =-A_{g}^{2} dt^{2}+B_{g}^{2} dr^{2}+r^{2} d\Omega^{2} \\
ds_{f}^{2} & =-A_{f}^{2} dt^{2}+B_{f}^{2} dr^{2}+2C_{f}^{2} dr dt+D_{f}^{2} d\Omega^{2}
\end{aligned}
\end{equation}
%%%%%
where $A_{g}, B_{g}, A_{f}, B_{f}, C_{f}$, and $D_{f}$ are the function of $r$ only. 
%We can exploit the gauge freedom to redefine  $D_{g}(r)=r$. 
Note that we have not made any prior assumptions about the metric $f_{\mu \nu}$.
%We as we see there is generally a cross-term $f_{r t}$ in the metric form of $f_{\mu \nu}$. 
All the components of $g_{\mu \nu}$ and $f_{\mu \nu}$ are determined by solving the field equation \autoref{einstein_equation}. 
%now we explore the field equation of $g_{\mu \nu}$ in the vacuum. 
%Here L.H.S is the Einstein Tensor $G_{\mu \nu}$ and $T_{\mu \nu}^{g}$ is the interaction term. as we stated in section II, we are dealing with the model in which three parameter $\beta_{0},\, \beta_{1}$ and $ \beta_{4}$ are non-zero\ref{}. other parameters $\beta_{2}$ and $\beta_{3}$  are taken Zero \ref{}. we have not assumed any restriction or specific value to the non-zero parameter. with this choice of parameter,  components of the L.H.S of the field equation \autoref{field_equation}.
The non-vanishing components of the Einstein tensor $G^{g}_{\mu \nu}$ for the metric $g_{\mu \nu}$ are given by
\begin{equation}
\begin{aligned}
& G^{g}_{00}=\frac{A_{g}^{2}}{r^{2}}\left[1-\left(\frac{r}{B_{g}^{2}}\right)^{\prime}\right]\,,\quad
 G^{g}_{11}=\frac{1}{r^{2}}-\frac{B_{g}^{2}}{r^{2}}+\frac{2 A_{g}^{\prime}}{r A_{g}}, \\
& G^{g}_{22}=\frac{r^{2}\left(-A_{g}^{\prime} B_{g}^{\prime}+B_{g} A_{g}^{\prime \prime}\right)+r\left(B_{g} A_{g}^{\prime}-A_{g} B_{g}^{\prime}\right)}{A_{g} B_{g}^{3}},\\
& G^{g}_{33}=\sin ^{2} \theta \; G_{22},\\
%& G_{\mu \nu} = 0\,  \;\;\;\;\;\forall \; \mu \neq \nu;
\end{aligned}
\end{equation}
Note that Einstein tensor is diagonal as $G^{g}_{\mu \nu}=0$  for $\mu \;\not=\; \nu$. Under the assumption $\beta_2=\beta_3=0$, the diagonal components of $T_{\mu\nu}^{g}$ are (see \autoref{effective_eom})
% under our model of only  $\beta_{0},\, \beta_{1}$ and $ \beta_{4}$ having non-zero. R.H.S of the field \autoref{field_equation} gives 
\begin{equation}
    \begin{aligned}
T_{00}^{g}&=-\frac{A_{g}^{2} m^{2}\left[\beta_{1} B_{f} r+B_{g}\left(2 \beta_{1} D_{f}+\beta_{0} r\right)\right]}{B_{g} r},\\
 T_{11}^{g}&=B_{g}^{2} m^{2}\left[\beta_{0}+\frac{\beta_{1} A_{f}}{A_{g}}+\frac{2 \beta_{1} D_{f}}{r}\right], \\
 T_{22}^{g}&=r \beta_{1} m^{2} D_{f}+r^{2} m^{2}\left[\beta_{0}+\frac{\beta_{1} A_{f}}{A_{g}}+\frac{\beta_{1} B_{f}}{B_{g}}\right], \\
T_{33}^{g}&=\sin ^{2} \theta T_{22}^{g},
\end{aligned}
\end{equation}
The non-vanishing off-diagonal components of the $T_{\mu \nu}^{g}$ are
\begin{equation}
\begin{aligned}
T_{10}^{g}=\frac{1}{2} i \beta_{1} m^{2}\left[A_{g}+i B_{g}\right] C_{f}, 
\end{aligned}
\end{equation}
and $T_{01}^{g}=T_{10}^{g}$. Since we are interested in vacuum solutions, the field equation \autoref{einstein_equation} dictates that $T_{01}^{g}$ must vanish which leads to the condition
\begin{equation}
\frac{1}{2}i\beta_{1} m^{2}\left(A_{g}+i B_{g}\right) C_{f}=0
\end{equation}
 In our model $\beta_{1} \neq 0$  and $ m \neq 0$. Thus, $
C_{f}$ should vanish identically to satisfy the above equation.
As a result, there are only  bi-diagonal  solutions if we take $\beta_{2}=\beta_{3}=0$ . It is worth noting that one can obtain  non-bidiagonal solutions by relaxing the above assumption \cite{Schmidt_May_2016}. 
  %%%%%%%%%%%%%%%%%%%%%%%%%%%%%%%%%%%%%%%%%%%%%%%%%%%%%%%%%%%%%%%%%
	%%%%%%%%%%%%%%%%%%%%%%%%%%%%%%%%%%%%%%%%%%%%%%%%%%%%%%%%%%%%%%%%%%%%%%%%%%%%%%%%%%%%%%%%%%%%%%%%%%%
	%%%%%%%%%%%%%%%%%%%%%%%%%%%%%%%%%%%%%%%%%%%%%%%%%%%%%%%%%%%%%%%%%%%%%%%%%%%%%%%%%%%%%%%%%%%%%%%
	%%%%%%%%%%%%%%%%%%%%%%%%%%%%%%%%%%%%%%%%%%%%%%%%%%%%%%%%%%%%%%%%%%%%%%%%%%%%%%%%%%%%%%%%%%%%%%%
	%%%%%%%%%%%%%%%%%%%%%%%%%%%%%%%%%%%%%%%%%%%%%%%%%%%%%%%%%%%%%%%%%%%%%%%%%%%%%%%%%%%%%%%%%%%%%%%%%%%
 
	\section{Roots of Cubic equation}\label{App:Circular}
 
 Consider the following cubic equation
  %%%%%%%%%%%%%%%%%%%%%%%%%%%%%%%%%%%%%%%%%%%%%%%%%%%%%%%%%%%%%%%%%
\begin{equation}\label{cubic_eq}
\begin{aligned}
r^3+a_2 r^2+a_1 r+a_0=0
\end{aligned}
\end{equation}
%%%%%%%%%%%%%%%%%%%%%%%%%%%%%%%%%%%%%%%%%%%%%%%%%%%%%%%%%%%%%%%%%
The roots of the above equation can be written as follows
  %%%%%%%%%%%%%%%%%%%%%%%%%%%%%%%%%%%%%%%%%%%%%%%%%%%%%%%%%%%%%%%%%
\begin{equation}\label{cubic_sol}
\begin{aligned}
r_1&=\left(S_1+S_2\right)-\frac{a_2}{2}\\
r_2&=-\frac{1}{2}\left(S_1+S_2\right)-\frac{a_2}{3}+\frac{i\sqrt{3}}{2}\left(S_1-S_2\right)\\
r_3&=-\frac{1}{2}\left(S_1+S_2\right)-\frac{a_2}{3}-\frac{i\sqrt{3}}{2}\left(S_1-S_2\right)
\end{aligned}
\end{equation}
%%%%%%%%%%%%%%%%%%%%%%%%%%%%%%%%%%%%%%%%%%%%%%%%%%%%%%%%%%%%%%%%%
\begin{widetext}
where

  %%%%%%%%%%%%%%%%%%%%%%%%%%%%%%%%%%%%%%%%%%%%%%%%%%%%%%%%%%%%%%%%%
\begin{equation}
\begin{aligned}
S_1&=\left[p+\sqrt{q^3+p^2}\right]^{1/3}\\
S_2&=\left[p-\sqrt{q^3+p^2}\right]^{1/3}
\end{aligned}
\end{equation}
%%%%%%%%%%%%%%%%%%%%%%%%%%%%%%%%%%%%%%%%%%%%%%%%%%%%%%%%%%%%%%%%%
and
%%%%%%%%%%%%%%%%%%%%%%%%%%%%%%%%%%%%%%%%%%%%%%%%%%%%%%%%%%%%%%%%%
\begin{equation}
\begin{aligned}
p&=\frac{1}{2}\left(a_1 a_2-3a_0\right)-\left(\frac{a_2}{3}\right)^3\\
q&=\left(\frac{a_1}{3}\right)-\left(\frac{a_2}{3}\right)^2
\end{aligned}
\end{equation}
%%%%%%%%%%%%%%%%%%%%%%%%%%%%%%%%%%%%%%%%%%%%%%%%%%%%%%%%%%%%%%%%%

Depending on the value of the $q^3+p^2$, we obtain three distinct types of roots; condition of which is given by the following expression
%%%%%%%%%%%%%%%%%%%%%%%%%%%%%%%%%%%%%%%%%%%%%%%%%%%%%%%%%%%%%%%%%

\begin{equation}
\begin{aligned}
q^3+p^2\begin{cases}
			>0, & \textrm{One real root and one pair of complex conjugate}\\
			=0, & \textrm{All real roots and at least two are equal}\\
   <0, & \textrm{All real roots }\\
		\end{cases}\\
\end{aligned}
\end{equation}

%%%%%%%%%%%%%%%%%%%%%%%%%%%%%%%%%%%%%%%%%%%%%%%%%%%%%%%%%%%%%%%%%
For \autoref{throat}, $q^3+p^2=(9M^2-1/\Lambda)/\Lambda^2$. Thus, the equation has three real roots if the following condition is satisfied $0<\Lambda\leq 1/9M^2$. Furthermore, in that scenario, $r_1=-(r_2+r_3)$.
\end{widetext}
%%%%%%%%%%%%%%%%%%%%%%%%%%%%%%%%%%%%%%%%%%%%%%%%%%%%%%%%%%%%%%%%%%%%%%%%%%%%%%%%%%%%%%%%%%%%%%%%%%%
	%%%%%%%%%%%%%%%%%%%%%%%%%%%%%%%%%%%%%%%%%%%%%%%%%%%%%%%%%%%%%%%%%%%%%%%%%%%%%%%%%%%%%%%%%%%%%%%%%%%

\bibliography{ref}

\bibliographystyle{utphys1}
	%%%%%%%%%%%%%%%%%%%%%%%%%%%%%%%%%%%%%%%%%%%%%%%%%%%%%%%%%%%%%%%%%%%%%%%%%%%%%%%%%%%%%%%%%%%%%%%%%%%
	%%%%%%%%%%%%%%%%%%%%%%%%%%%%%%%%%%%%%%%%%%%%%%%%%%%%%%%%%%%%%%%%%%%%%%%%%%%%%%%%%%%%%%%%%%%%%%%%%%%
	%%%%%%%%%%%%%%%%%%%%%%%%%%%%%%%%%%%%%%%%%%%%%%%%%%%%%%%%%%%%%%%%%%%%%%%%%%%%%%%%%%%%%%%%%%%%%%%%%%%
\end{document}